\documentclass[%
reprint,
superscriptaddress,
 amsmath,amssymb,
 aps, physrev,
]{revtex4-2}

\usepackage{color}
\usepackage{xcolor}
\usepackage{graphicx}
\usepackage{dcolumn}
\usepackage{siunitx}
\usepackage{bm}
\usepackage[version=4]{mhchem}
\usepackage{hyperref}
\hypersetup{colorlinks=true,linkcolor=blue,citecolor=blue,urlcolor=blue}

\usepackage[colorinlistoftodos]{todonotes}

\usepackage{soul} 

\begin{document}
\title{
Surface reconstruction and orthogonal decoupling in \ce{SrAl4} and \ce{EuAl4}
}

\author{Tongrui Li}
\thanks{Equal contributions}
\affiliation{National Synchrotron Radiation Laboratory, University of Science and Technology of China, Hefei, Anhui 230029, China}

\author{Leiyuan Chen}
\thanks{Equal contributions}
\affiliation{Laboratory of Advanced Materials, Fudan University, Shanghai, China}

\author{Jian Yuan}
\thanks{Equal contributions}
\affiliation{State Key Laboratory of Quantum Functional Materials, School of Physical Science and Technology, ShanghaiTech University, Shanghai 201210, China}
\affiliation{ShanghaiTech Laboratory for Topological Physics, ShanghaiTech University, Shanghai 201210, China}

\author{Zhengtai Liu}
\email{liuzt@sari.ac.cn}
\affiliation{Shanghai Synchrotron Radiation Facility, Shanghai Advanced Research Institute, Chinese Academy of Sciences, Shanghai 201210, China}
\affiliation{Shanghai Institute of Microsystem and Information Technology, Chinese Academy of Sciences, Shanghai 200050, China}

\author{Yichen Yang}
\affiliation{Shanghai Synchrotron Radiation Facility, Shanghai Advanced Research Institute, Chinese Academy of Sciences, Shanghai 201210, China}
\affiliation{Shanghai Institute of Microsystem and Information Technology, Chinese Academy of Sciences, Shanghai 200050, China}

\author{Zhicheng Jiang}
\affiliation{National Synchrotron Radiation Laboratory, University of Science and Technology of China, Hefei, Anhui 230029, China}

\author{Jianyang Ding}
\affiliation{Shanghai Synchrotron Radiation Facility, Shanghai Advanced Research Institute, Chinese Academy of Sciences, Shanghai 201210, China}
\affiliation{Shanghai Institute of Microsystem and Information Technology, Chinese Academy of Sciences, Shanghai 200050, China}

\author{Jiayu Liu}
\affiliation{State Key Laboratory of Functional Materials for Informatics, Shanghai Institute of Microsystem and Information Technology, Chinese Academy of Sciences, Shanghai 200050, China}

\author{Jishan Liu}
\affiliation{Shanghai Synchrotron Radiation Facility, Shanghai Advanced Research Institute, Chinese Academy of Sciences, Shanghai 201210, China}
\affiliation{Shanghai Institute of Microsystem and Information Technology, Chinese Academy of Sciences, Shanghai 200050, China}

\author{Zhe Sun}
\affiliation{National Synchrotron Radiation Laboratory, University of Science and Technology of China, Hefei, Anhui 230029, China}

\author{Yanfeng Guo}
\email{guoyf@shanghaitech.edu.cn}
\affiliation{State Key Laboratory of Quantum Functional Materials, School of Physical Science and Technology, ShanghaiTech University, Shanghai 201210, China}
\affiliation{ShanghaiTech Laboratory for Topological Physics, ShanghaiTech University, Shanghai 201210, China}

\author{Tong Zhang}
\email{Tzhang18@fudan.edu.cn}
\affiliation{Laboratory of Advanced Materials, Fudan University, Shanghai, China}

\author{Dawei Shen}
\email{dwshen@ustc.edu.cn}
\affiliation{National Synchrotron Radiation Laboratory, University of Science and Technology of China, Hefei, Anhui 230029, China}

\begin{abstract}

Surface-induced symmetry breaking in quantum materials can stabilize exotic electronic phases distinct from those in the bulk, yet its momentum-space manifestations remain elusive due to the domain averaging effects. Here, by utilizing angle-resolved photoemission spectroscopy (ARPES) and scanning tunneling microscopy (STM), we present a microscopic investigation of the electronic structures of \ce{SrAl4} and \ce{EuAl4}—layered tetragonal intermetallic compounds that exhibit well-characterized incommensurate charge density wave (CDW) transitions. Below the CDW transition temperatures, we uncover linearly dispersing electronic states and pronounced unidirectional replica bands orthogonal to the bulk CDW wave vector, evidencing the emergence of an in-plane $C_4$ symmetry-breaking electronic order that is not dictated by the bulk incommensurate CDW. STM measurements further reveal a 1$\times$2 surface reconstruction with quasi-one-dimensional modulations and half-unit-cell steps, traced to ordered 50\% Sr/Eu vacancies, which vanish irreversibly upon thermal cycling, indicating decoupled surface and bulk orders. These findings establish \ce{SrAl4} and \ce{EuAl4} as model platforms for exploring surface-confined nematicity and emergent low-dimensional phases in quantum materials.

\end{abstract}

\maketitle
\clearpage
\section{Introduction}

\begin{figure*}[htpb]
\centerline{\includegraphics[width=16.25cm]{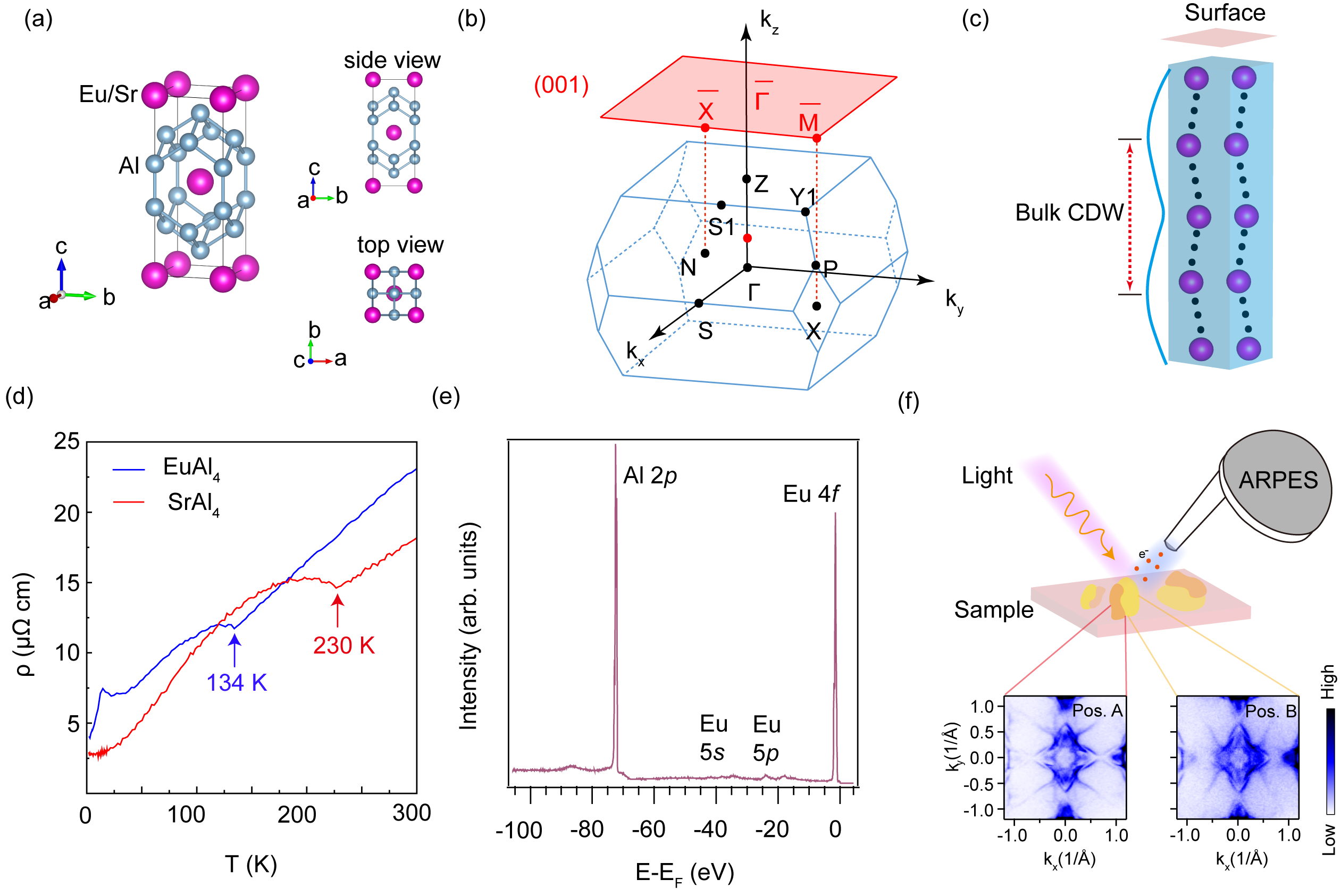}}
\caption{
\textbf{a} Crystal structures of \ce{SrAl4} and \ce{EuAl4}, shown along three principal axes;
\textbf{b} Three-dimensional BZ in reciprocal space, with the red shaded region indicating the two-dimensional projection onto the (001) plane;
\textbf{c} A schematic illustration depicting the propagation of the incommensurate CDW in the bulk;
\textbf{d} Temperature-dependent resistivity curves for \ce{EuAl4} and \ce{SrAl4};
\textbf{e} Core-level peak positions derived from integrated X-ray photoelectron spectroscopy (XPS) measurements using synchrotron radiation;
\textbf{f}  Fermi surfaces of \ce{SrAl4} measured at different positions at 12 K.
}
\label{Fig 1}
\end{figure*}

Electronic nematicity, characterized by spontaneous breaking of rotational symmetry while usually preserving translational symmetry, has emerged as a central focus in the study of quantum materials. It was discovered to be intimately linked to key unresolved questions in condensed matter physics, including the mechanisms driving high-temperature superconductivity in cuprates \cite{Lawler2010,Wu2017} and iron-based high-temperature superconductors \cite{Böhmer2022,227003,157002,aab0103}, as well as unconventional CDW order \cite{Asaba2024,Yang2024,Jiang2023}. Notably, recent experiments have extended the investigation of electronic nematicity from bulk to the surfaces and interfaces of materials \cite{021018,2308972120}. Within these confined low-dimensional systems, enhanced symmetry breaking is particularly pronounced and  can stabilize electronic phases that are absent in the bulk \cite{Du2021,5194,Wang2022222,Yuan2019}, offering a unique platform to explore the underlying physics of electronic nematicity.

While bulk nematicity has been extensively studied, surface-induced nematicity, where surface defects such as vacancies or reconstructions trigger rotation symmetry breaking, remains comparatively poorly understood, particularly in CDW materials. Currently, experiments on surface-induced nematicity have been primarily limited to spatially resolved techniques, leaving the reciprocal-space signatures of nematic transition largely unexplored \cite{1181083}. A major challenge in addressing this gap is the issue of domain averaging: Surface-induced nematicity typically emerges in micro/nanoscale domains with orthogonal orientations, accompanied by spontaneously formed twinned structures. Consequently, momentum-resolved techniques such as ARPES are inherently susceptible to such domain averaging effects, which obscure the detection of anisotropic electronic features and hinder the identification of nematic electronic structures in momentum space.

To overcome these limitations and gain deeper insights into surface-driven rotation symmetry breaking, it is crucial to identify material platforms that combine well-characterized bulk CDW transitions with structural simplicity and clean surfaces. The \ce{AB4} intermetallic compounds (A = Sr, Eu, Ba; B = Al, Ga), crystallizing in the tetragonal $I4/mmm$ structure, provide an ideal platform for such investigations. Despite their chemical simplicity, these materials exhibit a rich diversity of magnetic, electronic, and topological properties~\cite{L020405,064436,Lei2023,Wang2021,Takagi2022,011053,Vibhakar2024}. Among them, \ce{SrAl4} and \ce{EuAl4} stand out for exhibiting incommensurate CDW transitions at 243 K and 140 K, respectively, both with wave vectors propagating along the crystallographic $c$-axis \cite{015001,124711,094703,014602,064704,023277}. Their bulk CDW orders have been thoroughly characterized using synchrotron X-ray diffraction \cite{045102,195150,023277}, Raman spectroscopy \cite{Cao2025}, and theoretical calculations \cite{Wang2024}. However, the surface electronic structure of these compounds, in particular together with their potential to host surface-specific symmetry-breaking phases such as nematicity, remains unexplored. The combination of a simple tetragonal lattice, layered structure, and well-understood bulk properties positions the \ce{AB4} family as a compelling platform for investigating the interplay between electronic structure, lattice dynamics, and symmetry breaking at the surface, where the reduced dimensionality may stabilize exotic electronic orders.

In this work, we synthesize high-quality single crystals of \ce{SrAl4} and \ce{EuAl4} and investigate their surface and bulk electronic structures using $\mu$-ARPES, STM and density functional theory (DFT). Below the CDW transition temperatures, we observe linearly dispersing electronic states near the Fermi level, accompanied by prominent, unidirectional replica bands that are orthogonal to the bulk CDW wave vector. These replica features indicate a breaking of intrinsic fourfold ($C_4$) rotational symmetry in the $a$–$b$ plane, suggestive of a surface electronic nematic order. STM measurements further reveal a 1$\times$2 surface reconstruction characterized by quasi-one-dimensional modulations and half-unit-cell height steps along the $c$-axis. First-principles calculations attribute this reconstruction to ~50\% surface Sr/Eu vacancy ordering, which we propose as a key stabilizing factor for the observed symmetry breaking. Notably, this surface order undergoes irreversible suppression upon thermal cycling, which further highlights its distinction from the bulk CDW state. Our findings unveil an unexpected, orthogonal decoupling between surface and bulk symmetry-breaking orders in \ce{SrAl4} and \ce{EuAl4}. These materials thus constitute model systems for studying surface-confined nematicity and other emergent low-dimensional electronic phases in quantum materials.

~\\

\noindent\textbf{Results}

\vspace{1em}

\noindent\textbf{Crystal and transport nature of \ce{SrAl4} and \ce{EuAl4} single crystals.} 

The compound \ce{EuAl4} and \ce{SrAl4} crystallize in the tetragonal $I4/mmm$ space group with lattice parameters $a$ = $b$ = 4.401 Å and $c$ = 11.164 Å for EuAl$_4$, and $a$ = $b$ = 4.450 Å and $c$ = 11.187 Å for SrAl$_4$ (Fig.1(a)). Each Eu or Sr atom is coordinated by sixteen Al atoms, forming eight shorter and eight longer Eu/Sr–Al bonds. This structure comprises two crystallographically distinct Al sites. {Al1} coordinated by four equivalent Eu or Sr atoms and eight Al atoms, forming a distorted cuboctahedron via corner/edge/face-sharing. {Al2} bonded to four Eu/Sr atoms and five Al atoms, yielding a nine-coordinate polyhedral geometry. The Al sublattice displays an eaves-like structural motif along the $a$-axis and $C_4$ symmetry along the $c$-axis. The corresponding Brillouin zones (BZ) of \ce{EuAl4} or \ce{SrAl4} adopt a flattened, truncated-octahedral geometry, and its (001) projection, highlighted in red in Fig.1(b), shows clear $C_{4}$ symmetry. 
The incommensurate CDW modulation direction ($q_{CDW}$ $\parallel [001]$) is illustrated in Fig.~1(c), with small $q_{CDW}$ complicating direct surface observation. Our transport measurements reveal resistivity anomalies at $\sim 230$~K (\ce{SrAl4}) and $\sim 134$~K (\ce{EuAl4}), consistent with bulk CDW transitions (Fig. 1(d)) \cite{124711,023277}. X-ray photoelectron spectroscopy (XPS) of \ce{EuAl4} (Fig. 1(e)) shows prominent Eu 4$f$ and Al 2$p$ peaks, with weaker  Eu 5$s$ and 5$p$ contributions. Hereafter, we focus on the band dispersion of \ce{EuAl4} and \ce{SrAl4} within 1 eV of the Fermi level.


\vspace{1em}

\noindent\textbf{Linear band dispersion on high-symmetry path of \ce{EuAl4}}

\begin{figure}[htp]
\centerline{\includegraphics[width=7.532cm]{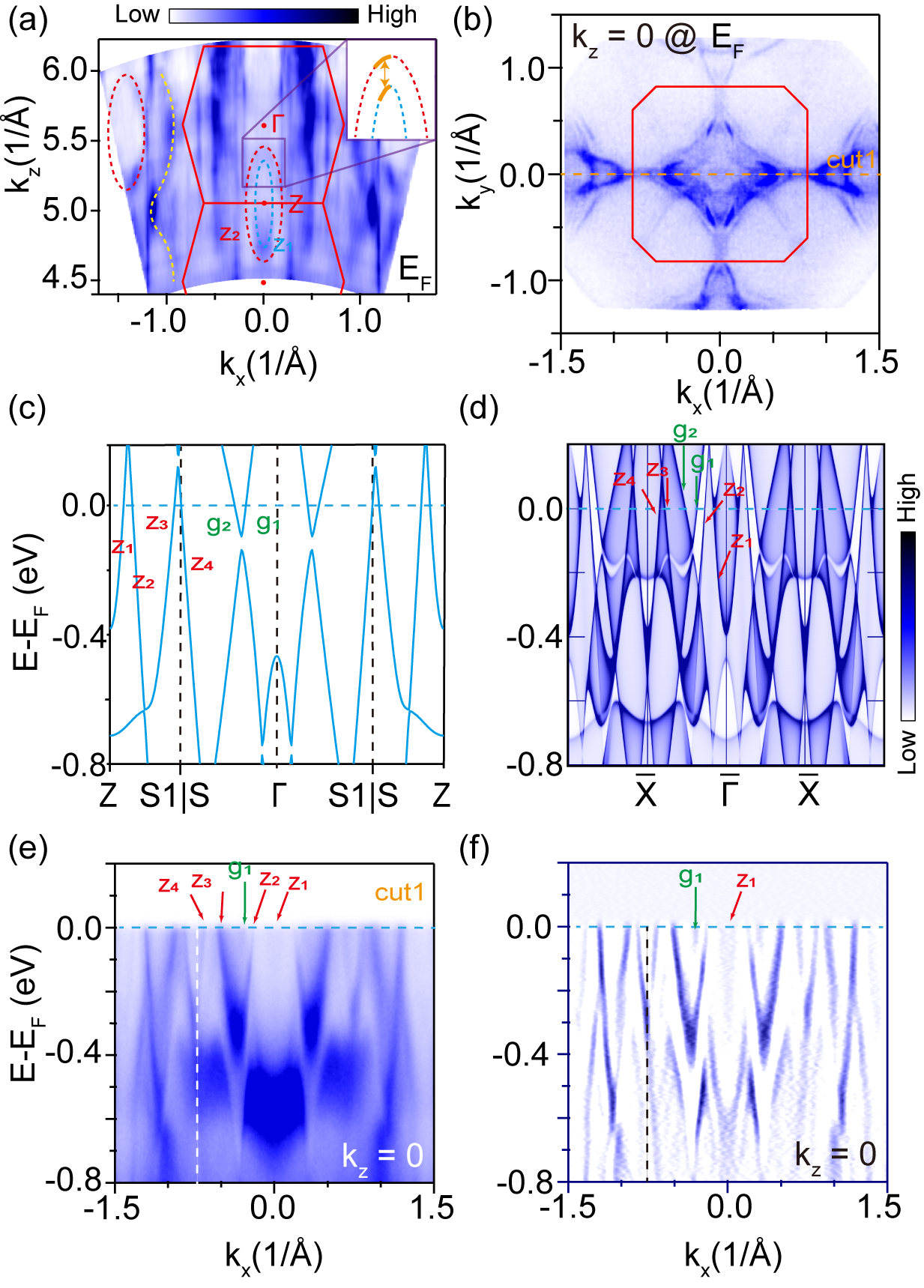}}
\caption{
\textbf{a} Fermi surface of \ce{EuAl4} in the $k_{x}$-$k_{z}$ maps measured by $\mu$-ARPES (The insets are the zoom-in view of the Fermi surface nesting vector) at 8 K;
\textbf{b} Fermi surface of \ce{EuAl4} in the $k_{x}$-$k_{y}$ maps;
\textbf{c} High-symmetry path calculated using DFT;
\textbf{d} Spectral function of the bulk projection along the $\overline{X}$-$\overline{\Gamma}$-$\overline{X}$ direction in the two-dimensional projected BZ;
\textbf{e}  ARPES intensity along the cut1 direction in the two-dimensional projected BZ;
\textbf{f} Second-derivative spectrum of (e).
}
\label{Fig 2}
\end{figure}

We first conducted spatially resolved $\mu$-ARPES scan across the entire cleaved  (001) surface of \ce{SrAl4} using a micro-focused beam. Strikingly, measurements at two distinct locations revealed Fermi surfaces with $C_2$ symmetry, deviating from the bulk $C_4$ symmetry, as shown in Fig. 1(f). 
By comparison, low-energy electron diffraction (LEED) on the same surface displays preserved $C_4$ symmetry (Supplementary Fig. S1), consistent with the (001) projection of BZ. Given the millimeter-scale LEED beam spot, substantially larger than that of $\mu$-ARPES, these results suggest that the origin of the unidirectional replica bands is linked to microscale domain effects.

To further elucidate the observed symmetry breaking in the electronic structure, it was first necessary to accurately determine the correspondence between photon energies and high-symmetry planes in momentum space within a single domain. Synchrotron-based ARPES measurements were therefore performed using photon energies ranging from 70 to 140 eV. As shown in Fig. 2(a), the constant-energy maps in the $k_x$–$k_z$ plane clearly reveal three-dimensional characteristics of the electronic structure in \ce{EuAl4} and well-defined periodicity along $k_z$, particularly within the second BZ (see Supplementary Fig. S2 for additional details). 
A periodic yellow band in Fig. 2(a) serves as a guide for accurate mapping of the band structure onto the BZ. Using the lattice constants of \ce{EuAl4} and an assumed inner potential of 12 eV, we extract a consistent periodicity in $k_z$. Specifically, photon energies of 90 eV and 138 eV correspond to the $k_z = \pi$ plane, while 69 eV and 112 eV map to the $k_z = 0$ plane. This periodicity remains robust even at deeper binding energies, as illustrated in Supplementary Fig. S2. Combining DFT calculations with ARPES measurements, the Fermi surface in the $k_x$–$k_z$ plane reveals two concentric elliptical pockets centered at the $Z$ point, labeled z1 and z2, with a nesting vector of approximately 0.108 Å$^{-1}$, as indicated by the orange arrow. This closely matches the bulk CDW wave vector $q_{CDW}$ 
= 0.1013 Å$^{-1}$. Despite this favorable nesting condition, no evidence of a CDW gap or Fermi surface reconstruction is observed in our ARPES spectra. This result offers compelling experimental evidence that the CDW order in this material arises from a mechanism beyond conventional Fermi surface nesting, which is consistent with Kobata \textit{et al.} \cite{094703} and the recent theoretical predictions by Wang \textit{et al.} \cite{Wang2024}. 

\begin{figure*}[htp!]
\centerline{\includegraphics[width=17cm]{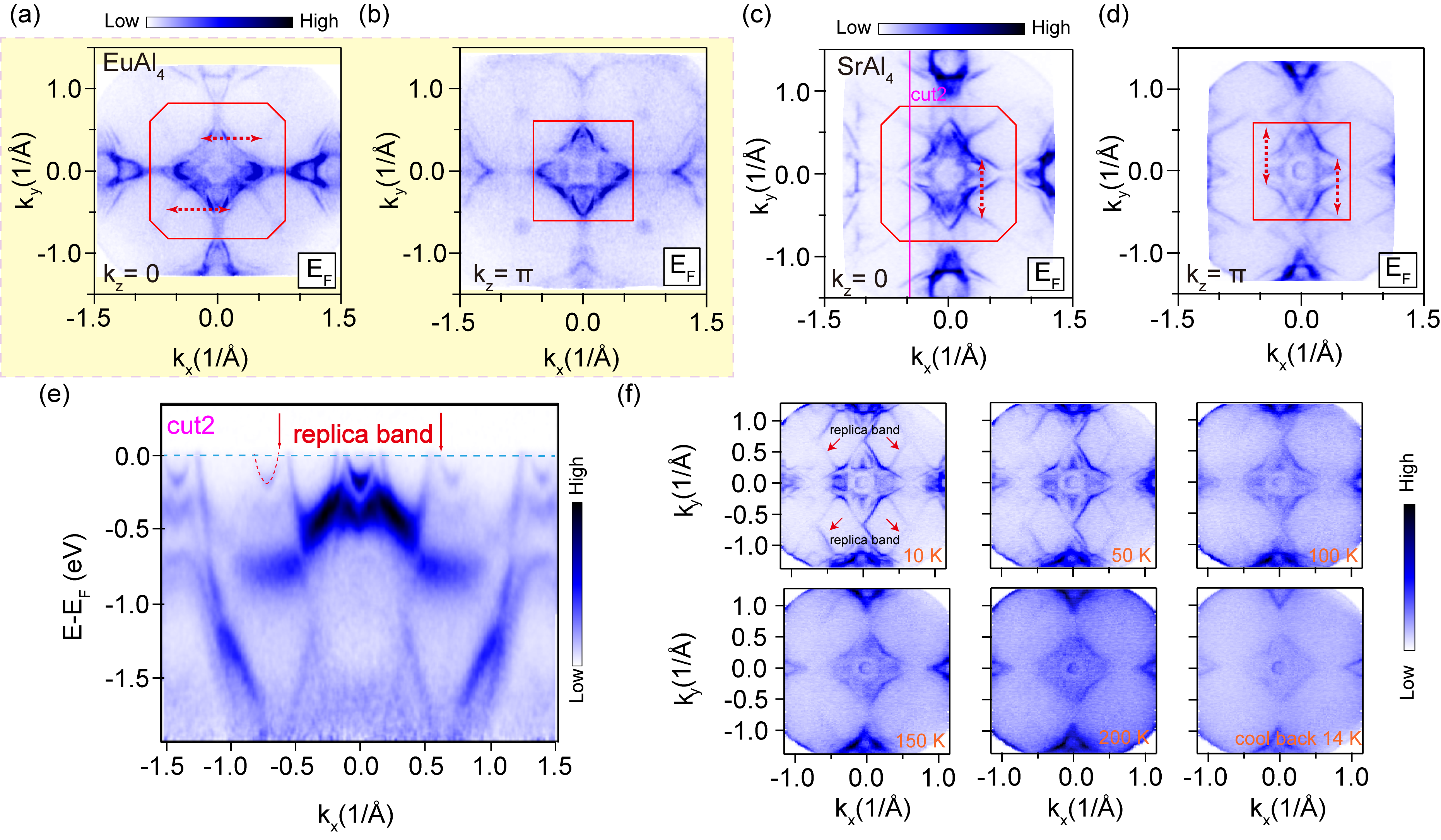}}
\caption{
\textbf{a b} Fermi surfaces of \ce{EuAl4} at $k_z = 0$ and $k_z = \pi$, respectively. The red arrow indicates the replication vector of the replication band;
\textbf{c d} Fermi surfaces of \ce{SrAl4} at $k_z = 0$ and $k_z = \pi$, respectively;
\textbf{e} ARPES intensity along the cut2 of \ce{SrAl4} in c;
\textbf{f} Temperature evolution of the Fermi surface of \ce{SrAl4}, with the replica band gradually weakening and not recovering after cooling.
}
\label{Fig 3}
\end{figure*}

Figure~2(b) presents the measured Fermi surface of \ce{EuAl4} in the $k_z = 0$ plane, where the BZ boundaries are clearly delineated by the red outline. As shown in Supplementary Fig. S3, the experimentally observed constant energy contours are in excellent agreement with the bulk electronic structure projections calculated via the surface Green's function method. The DFT calculations, as shown in Fig. 2(c), further reveal the presence of six bands crossing the Fermi level along the high-symmetry path $\Gamma$–S(S$_1$)–Z, labeled as $g_1$, $g_2$, $z_1$, $z_2$, $z_3$, and $z_4$. These bands exhibit pronounced linear dispersion, consistent with the electronic structure previously reported ~\cite{094703}. Figures~2(e) and 2(f) display ARPES and corresponding second-derivative spectrum, respectively, measured along Z–S direction (cut\#1). Notably, due to the body-centered tetragonal lattice structure of \ce{EuAl4}, there exists a half reciprocal lattice vector shift between the first and second Brillouin zones along the $k_z$ direction (Fig. 1b). This leads to the $\Gamma$ point of the first BZ and the $Z$ point of the second BZ residing within the same momentum plane, enabling their simultaneous detection by photons of identical energy. Additionally, although the employed photon energy corresponds to the $k_z = 0$ plane, the ARPES spectra not only resolve the $g_1/g_2$ bands but also successfully capture the $z_1$–$z_4$ bands located within the $k_z = \pi$ plane along the Z–S$_1$ path. This observation probably originates from the $k_z$-broadening effect, which is quantitatively reproduced through the calculation of the two-dimensional projected spectral function along the $\bar\Gamma$–$\bar{X}$ path (Fig. 2(d)).

\vspace{1em}
\noindent\textbf{Unidirectional replica bands in \ce{EuAl4}/\ce{SrAl4}}

\begin{figure*}[htp]
\centerline{\includegraphics[width=17.1cm]{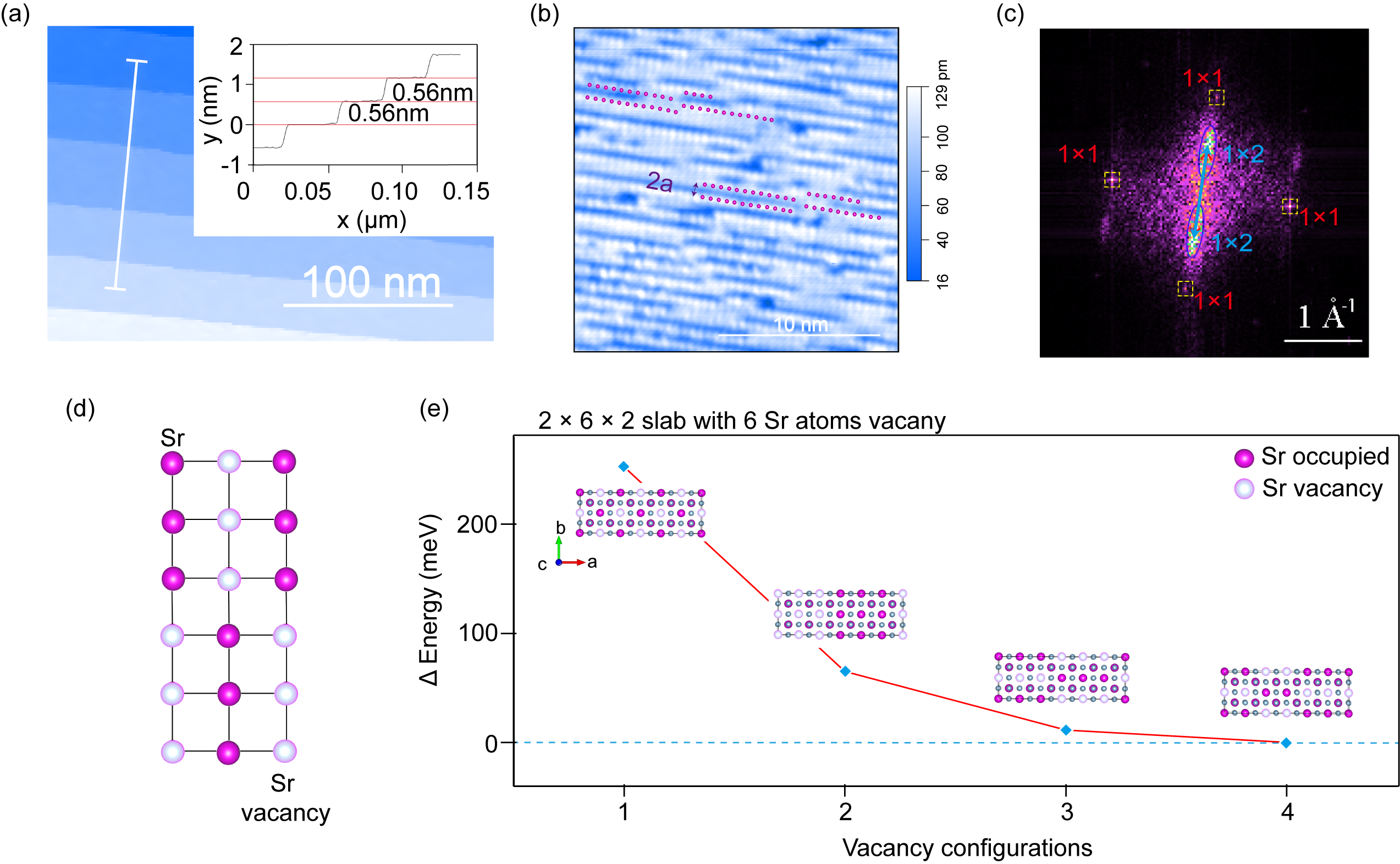}}
\caption{
\textbf{a}  Surface step morphology of \ce{SrAl4}, with the inset showing the step height;
\textbf{b} Atomic resolution and surface reconstruction measured by STM (bias voltage $V_{b}$ = 20 mV, tunneling current $I_{t}$ = 5 nA, measured temperature at 4 K);
\textbf{c} Fast Fourier transform (FFT) analysis of the surface morphology in (b), accompanied by pronounced $1 \times 2$ peaks; 
\textbf{d} Schematic of the 1×2 surface reconstruction, where Sr atoms tend to form alternating one-dimensional chains due to the presence of vacancies;
\textbf{e} Total energy difference profiles for different Sr-vacancy configurations calculated using $2 \times 6 \times 2$ slab models, showing that the ground state with alternating one-dimensional Sr chains is energetically favored.
}
\label{Fig 4}
\end{figure*}

Although DFT calculations largely reproduce the band dispersion and Fermi surface morphology observed in ARPES experiments, they fail to reasonably explain the in-plane symmetry breaking from $C_4$ to $C_2$ observed on the Fermi surface. To delve into the origin of this symmetry breaking, we systematically investigated the electronic structures of \ce{EuAl4} and \ce{SrAl4} across the three-dimensional BZ, as well as their temperature-dependent evolution. Figures 3(a–d) respectively display the Fermi surface topology of \ce{EuAl4} and \ce{SrAl4} at the $k_z=0$ and $k_z=\pi$ planes under low temperatures. In addition to bulk bands, anomalous band branches along a single crystalline axis direction are observed, directly evidencing in-plane $C_4$ symmetry breaking. Notably, the identification of these anomalous bands in \ce{SrAl4} is significantly more pronounced than in \ce{EuAl4}, which could be attributable to its higher CDW transition temperature and flatter cleaved surface.

Further insight is provided by the photoemission intensity plot shown in Figure 3(e), which is measured along a non-high-symmetry direction marked in Figure 3(c). This reveals a strikingly similar dispersion between the anomalous band and the main band in the first BZ. Such a feature naturally suggests the possibility of CDW-induced band replication. However, it should be noted that previous studies confirm that the bulk CDW wave vector in this system is oriented exclusively along the $c$-axis. Consequently, the observed in-plane replica bands should originate from a novel physical mechanism independent of the bulk CDW order. Furthermore, as demonstrated in Figure 3(f), the spectral intensity of these replica bands diminishes significantly with increasing temperature and cannot be recover upon subsequent cooling. This irreversible temperature-dependent behavior stands in stark contrast to the typical thermodynamic response of bulk electronic orders, strongly supporting a scenario dominated by surface reconstruction or defect-induced symmetry breaking.

Given the pronounced surface sensitivity of ARPES within the measured photon energy range, complementary surface-sensitive techniques are essential to elucidate the microscopic origin of replica bands. STM performed at low temperatures on freshly cleaved \ce{SrAl4} surfaces, reveals atomically resolved surface features that provide crucial insight. As shown in Fig. 4a, STM topographs display distinct step edge with a height of approximately 0.56 nm, corresponding to half the  $c$-axis lattice constant, which is consistent with a cleavage plane located between Eu and Al layers. Theoretical calculations of exfoliation energy \cite{Jung2018} identify two bond-breaking modes along the (001) direction of \ce{EuAl4}—Eu–Al and Al–Al. The Eu–Al bond exhibits a significantly lower exfoliation energy (Supplementary Fig. S4), further supporting the experimentally inferred cleavage pathway. Atomic-resolution STM images (Fig. 4b) reveal a pronounced unidirectional modulation, suggesting the emergence of a quasi-one-dimensional electronic state aligned along a specific in-plane direction. This modulation manifests as parallel chains separated by one atomic layer (with a spacing of 2$a$).
The corresponding fast fourier transform (FFT) pattern displays a well-defined 1×2 periodicity (Fig. 4c), confirming the unidirectional structural reconstruction.  

While this behavior differs from that reported in \ce{EuTe4} \cite{Xiao2024}, it closely resembles phenomena observed in other Eu-based compounds, such as \ce{EuIn2As2} \cite{Li2021}. We attribute the observed 1 × 2 surface reconstruction to a spontaneous ordering driven by 50\% coverage of Eu or Sr atoms, as illustrated in Fig. 4d. The calculations presented in Fig.\thinspace 4(e) (see details in Fig.\thinspace S5), performed in the absence of bulk CDW order, demonstrate that configurations with ordered Sr vacancies (Configs. 3 and 4) exhibit a ground-state energy reduction exceeding 0.2 eV compared to other arrangements. 
These energetically favorable vacancy configurations precisely reproduce the $1 \times 2$ periodicity observed in the STM topographs (Fig.~4b), where quasi-1D chain-like vacancy arrangements alternate between surface layers.

Further analysis of the correlation length along the vertical chain-like direction (see details in Fig.S6) reveals a short-range order of approximately 0.49 nm, indicating that the 1D chain-like features tend to alternate between layers. Upon undergoing a thermal cycle (i.e., heating followed by cooling), the 1 × 2 electronic modulation becomes undetectable in STM measurements. This irreversibility aligns with the disappearance of replica bands, confirming the metastable nature of the defect-driven surface reconstruction. 
These findings collectively highlight a clear thermodynamic origin for the surface reconstruction observed in \ce{SrAl4}, driven by energetically favored vacancy ordering. The metastability and irreversibility of the reconstructed state underscore the delicate balance between cleavage dynamics, surface defect formation, and electronic symmetry breaking.

~\\
\noindent\textbf{Discussion}

In summary, we have revealed a previously unrecognized, surface-confined electronic symmetry breaking in the prototypical layered compounds \ce{SrAl4} and \ce{EuAl4}. Using the combination of $\mu$-ARPES, STM and DFT calculations, we have identified a robust $C_4 \rightarrow C_2$ symmetry breaking on the (001) surfaces of both materials, which manifests as unidirectional replica bands in momentum space and a quasi-one-dimensional $1 \times 2$ surface reconstruction in real space. Crucially, the emergence of these in-plane replica bands—clearly resolved in both SrAl$_4$ and EuAl$_4$—cannot be reconciled with the bulk incommensurate CDW order, which propagates strictly along the crystallographic $c$-axis. This orthogonality between the in-plane replica vectors and the out-of-plane CDW wave vector indicates a symmetry-breaking mechanism that is decoupled from the bulk and confined to the surface. This stands in stark contrast to known cases in Fe-based superconductors~\cite{227003,157002,aab0103}, \ce{Sr2RuO4}~\cite{5194}, and conventional CDW systems \cite{5068362,205101,5118870,Liu2024,1424791112,Chen2015}, where surface and bulk symmetry breaking are typically aligned or directly coupled. Our STM measurements provide critical insights into the microscopic origin of this surface-specific ordering. The $1 \times 2$ reconstruction, revealed through both topographic imaging and Fourier-transform analysis, is attributed to a spontaneous ordering of 50\% Sr or Eu vacancies on the surface layer. This scenario is corroborated by calculations, which reproduce the observed superstructure. Moreover, the disappearance of both the replica bands and the $1 \times 2$ modulation upon thermal cycling underscores the metastable nature of this surface order and its detachment from the bulk phase diagram.

Our work establish SrAl$_4$ and EuAl$_4$ as model systems in which surface symmetry breaking emerges independently of bulk instabilities. This orthogonal decoupling between surface and bulk electronic orders represents a notable deviation from traditional paradigms in layered quantum materials, where surface and bulk transitions are often intertwined. Our findings open new avenues for tailoring emergent low-dimensional phases via surface engineering, such as controlled vacancy ordering or heterointerface design. More broadly, this work offers a compelling platform for exploring the interplay between surface and bulk symmetry breaking, and suggests potential strategies for engineering novel surface-confined quantum states through targeted structural modulation.

~\\
\noindent\textbf{Methods}

\noindent\textbf{Sample growth and characterization}

The EuAl$_{4}$ and SrAl$_{4}$ single crystals were grown by using self-flux method. Starting materials of Eu/Sr and Al were mixed in a molar ration of 1:10 and placed into an alumina crucible which was then sealed into a quartz tube in vacuum. The assembly was heated in a furnace up to 950 °C within 10 h, kept at that temperature for 20 h, and then slowly cooled down to 700 °C at a temperature decreasing rate of 2 °C/h. The excess Al was removed at this temperature by quickly placing the assembly into a high-speed centrifuge. The phase and quality of the crystals were examined on the Bruker D8 single crystal x-ray diffractometer with Mo K$\alpha_1$ ($\lambda$ = 0.71073 Å).

\noindent\textbf{ARPES experiments}

All \ce{EuAl4} and \ce{SrAl4} single crystal samples were cleaved {\it in situ} at 7\,K with a base pressure of better than 6 $\times$ 10$^{-11}$\,Torr. High-resolution ARPES measurements were performed at the 03U beamline of Shanghai Synchrotron Radiation Facility (SSRF)\cite{yang2021high} with liner horizontal polarization lights~\cite{sun2020performance}. In our measurements. light's linear horizontal polarization is parallel to the ground and the incident angle on sample is $\theta$ = 45$^{\circ}$ with respect to the sample's normal direction, along with the slit direction that is perpendicular to the ground. All data were acquired with a Scienta-Omicron DA30 electron analyzer. The total energy resolution was set to 10$\sim$20\,meV depending on the photon energy applied, and the angular resolution was set to be 0.2$^{\circ}$. In our experiments, the photon energy region is from 70 to 140 eV, and the inner potential of \ce{EuAl4} and \ce{SrAl4} determined by our photon energy dependent measurement is about 12 eV.

\noindent\textbf{Band calculations}

The electronic structure of \ce{EuAl4} and \ce{SrAl4} was calculated using density functional theory (DFT) within the projector augmented-wave (PAW) method \cite{prb5017}, as implemented in the Vienna ab initio Simulation Package (VASP) \cite{prb511169}. The exchange–correlation functional was treated within the generalized gradient approximation (GGA) parameterized by Perdew, Burke, and Ernzerhof (PBE) \cite{prl773865}. Brillouin-zone integrations were performed on an $8 \times 8 \times 11$ Monkhorst–Pack $k$-point mesh, with a plane-wave energy cutoff of 400 eV. Experimental lattice parameters were used, and spin–orbit coupling (SOC) was included in all calculations. To reproduce the spectral functions in the paramagnetic state, a tight-binding model based on maximally localized Wannier functions \cite{prb5612847,prb65035109}, which included Eu $s$, $p$ and Al $s$, $p$ orbitals. In the calculations, the Fermi level was shifted downward by 0.25 eV to match the experimental results. Two-dimensional Fermi surfaces were calculated using iterative Green function methods \cite{mpls,mpls2} as implemented in WANNIERTOOLS \cite{wuwt}.

The authors declare that the main data supporting the findings of this study are available within the paper and its Supplementary Material. Extra data are available from the corresponding authors upon request.

\noindent\textbf{Acknowledgements}

This work is supported by National Key R\&D Program of China (Grants No. 2023YFA1406304 and 2024YFA1408103), National Science Foundation of China (Grants No. 12494593, and 12004405) and Anhui Provincial Natural Science Foundation (No. 2408085J003). Y. F. G was sponsored by Double First-Class Initiative Fund of ShanghaiTech University and Beijing National Laboratory for Condensed Matter Physics (2023BNLCMPKF002). J.M. was supported by the National Key Research and Development Program of China (Grant No. 2022YFA1402704). Part of this research used Beamline 03U of the Shanghai Synchrotron Radiation Facility, which is supported by ME$^2$ project under Contract No.11227902 from National Natural Science Foundation of China. The authors also thank the support from Analytical Instrumentation Center (\#SPST-AIC10112914).
.

\noindent\textbf{Author contributions}

T.L., Z.J., Y.Y., Z.T. and D.S. performed the ARPES experiment and analyzed the resulting data. L.C. and T.Z. performed the STM experiment. T.L. and Z.S. performed the theoretical calculations. J.Y. and Y.G. synthesized and characterized the single crystals. Z.L., T.Z., Y.G., D.S. supervised the project. Z.J., J.D., J.L., J.L. and Z.L. contributed to measurements and data analysis. T.L., L.C., J.Y., Z.T. and D.S. wrote the manuscript with input from all coauthors.

\noindent\textbf{Competing interests}

The authors declare no competing interests.

\bibliography{main}

\begin{thebibliography}{52}%
\makeatletter
\providecommand \@ifxundefined [1]{%
 \@ifx{#1\undefined}
}%
\providecommand \@ifnum [1]{%
 \ifnum #1\expandafter \@firstoftwo
 \else \expandafter \@secondoftwo
 \fi
}%
\providecommand \@ifx [1]{%
 \ifx #1\expandafter \@firstoftwo
 \else \expandafter \@secondoftwo
 \fi
}%
\providecommand \natexlab [1]{#1}%
\providecommand \enquote  [1]{``#1''}%
\providecommand \bibnamefont  [1]{#1}%
\providecommand \bibfnamefont [1]{#1}%
\providecommand \citenamefont [1]{#1}%
\providecommand \href@noop [0]{\@secondoftwo}%
\providecommand \href [0]{\begingroup \@sanitize@url \@href}%
\providecommand \@href[1]{\@@startlink{#1}\@@href}%
\providecommand \@@href[1]{\endgroup#1\@@endlink}%
\providecommand \@sanitize@url [0]{\catcode `\\12\catcode `\$12\catcode `\&12\catcode `\#12\catcode `\^12\catcode `\_12\catcode `\%12\relax}%
\providecommand \@@startlink[1]{}%
\providecommand \@@endlink[0]{}%
\providecommand \url  [0]{\begingroup\@sanitize@url \@url }%
\providecommand \@url [1]{\endgroup\@href {#1}{\urlprefix }}%
\providecommand \urlprefix  [0]{URL }%
\providecommand \Eprint [0]{\href }%
\providecommand \doibase [0]{https://doi.org/}%
\providecommand \selectlanguage [0]{\@gobble}%
\providecommand \bibinfo  [0]{\@secondoftwo}%
\providecommand \bibfield  [0]{\@secondoftwo}%
\providecommand \translation [1]{[#1]}%
\providecommand \BibitemOpen [0]{}%
\providecommand \bibitemStop [0]{}%
\providecommand \bibitemNoStop [0]{.\EOS\space}%
\providecommand \EOS [0]{\spacefactor3000\relax}%
\providecommand \BibitemShut  [1]{\csname bibitem#1\endcsname}%
\let\auto@bib@innerbib\@empty
\bibitem [{\citenamefont {Lawler}\ \emph {et~al.}(2010)\citenamefont {Lawler}, \citenamefont {Fujita}, \citenamefont {Lee}, \citenamefont {Schmidt}, \citenamefont {Kohsaka}, \citenamefont {Kim}, \citenamefont {Eisaki}, \citenamefont {Uchida}, \citenamefont {Davis}, \citenamefont {Sethna},\ and\ \citenamefont {Kim}}]{Lawler2010}%
  \BibitemOpen
  \bibfield  {author} {\bibinfo {author} {\bibfnamefont {M.~J.}\ \bibnamefont {Lawler}}, \bibinfo {author} {\bibfnamefont {K.}~\bibnamefont {Fujita}}, \bibinfo {author} {\bibfnamefont {J.}~\bibnamefont {Lee}}, \bibinfo {author} {\bibfnamefont {A.~R.}\ \bibnamefont {Schmidt}}, \bibinfo {author} {\bibfnamefont {Y.}~\bibnamefont {Kohsaka}}, \bibinfo {author} {\bibfnamefont {C.~K.}\ \bibnamefont {Kim}}, \bibinfo {author} {\bibfnamefont {H.}~\bibnamefont {Eisaki}}, \bibinfo {author} {\bibfnamefont {S.}~\bibnamefont {Uchida}}, \bibinfo {author} {\bibfnamefont {J.~C.}\ \bibnamefont {Davis}}, \bibinfo {author} {\bibfnamefont {J.~P.}\ \bibnamefont {Sethna}},\ and\ \bibinfo {author} {\bibfnamefont {E.-A.}\ \bibnamefont {Kim}},\ }\bibfield  {title} {\bibinfo {title} {Intra-unit-cell electronic nematicity of the high-\ce{Tc} copper-oxide pseudogap states},\ }\href {https://doi.org/10.1038/nature09169} {\bibfield  {journal} {\bibinfo  {journal} {Nature}\ }\textbf {\bibinfo {volume} {466}},\ \bibinfo {pages} {347}
  (\bibinfo {year} {2010})}\BibitemShut {NoStop}%
\bibitem [{\citenamefont {Wu}\ \emph {et~al.}(2017{\natexlab{a}})\citenamefont {Wu}, \citenamefont {Bollinger}, \citenamefont {He},\ and\ \citenamefont {Bo{\v{z}}ovi{\'{c}}}}]{Wu2017}%
  \BibitemOpen
  \bibfield  {author} {\bibinfo {author} {\bibfnamefont {J.}~\bibnamefont {Wu}}, \bibinfo {author} {\bibfnamefont {A.~T.}\ \bibnamefont {Bollinger}}, \bibinfo {author} {\bibfnamefont {X.}~\bibnamefont {He}},\ and\ \bibinfo {author} {\bibfnamefont {I.}~\bibnamefont {Bo{\v{z}}ovi{\'{c}}}},\ }\bibfield  {title} {\bibinfo {title} {Spontaneous breaking of rotational symmetry in copper oxide superconductors},\ }\href {https://doi.org/10.1038/nature23290} {\bibfield  {journal} {\bibinfo  {journal} {Nature}\ }\textbf {\bibinfo {volume} {547}},\ \bibinfo {pages} {432} (\bibinfo {year} {2017}{\natexlab{a}})}\BibitemShut {NoStop}%
\bibitem [{\citenamefont {B{\"o}hmer}\ \emph {et~al.}(2022)\citenamefont {B{\"o}hmer}, \citenamefont {Chu}, \citenamefont {Lederer},\ and\ \citenamefont {Yi}}]{Böhmer2022}%
  \BibitemOpen
  \bibfield  {author} {\bibinfo {author} {\bibfnamefont {A.~E.}\ \bibnamefont {B{\"o}hmer}}, \bibinfo {author} {\bibfnamefont {J.-H.}\ \bibnamefont {Chu}}, \bibinfo {author} {\bibfnamefont {S.}~\bibnamefont {Lederer}},\ and\ \bibinfo {author} {\bibfnamefont {M.}~\bibnamefont {Yi}},\ }\bibfield  {title} {\bibinfo {title} {Nematicity and nematic fluctuations in iron-based superconductors},\ }\href {https://doi.org/10.1038/s41567-022-01833-3} {\bibfield  {journal} {\bibinfo  {journal} {Nature Physics}\ }\textbf {\bibinfo {volume} {18}},\ \bibinfo {pages} {1412} (\bibinfo {year} {2022})}\BibitemShut {NoStop}%
\bibitem [{\citenamefont {Zhang}\ \emph {et~al.}(2016)\citenamefont {Zhang}, \citenamefont {Park}, \citenamefont {Lu}, \citenamefont {Wei}, \citenamefont {Ma}, \citenamefont {Hao}, \citenamefont {Dai}, \citenamefont {Meng}, \citenamefont {Yang}, \citenamefont {Luo},\ and\ \citenamefont {Li}}]{227003}%
  \BibitemOpen
  \bibfield  {author} {\bibinfo {author} {\bibfnamefont {W.}~\bibnamefont {Zhang}}, \bibinfo {author} {\bibfnamefont {J.~T.}\ \bibnamefont {Park}}, \bibinfo {author} {\bibfnamefont {X.}~\bibnamefont {Lu}}, \bibinfo {author} {\bibfnamefont {Y.}~\bibnamefont {Wei}}, \bibinfo {author} {\bibfnamefont {X.}~\bibnamefont {Ma}}, \bibinfo {author} {\bibfnamefont {L.}~\bibnamefont {Hao}}, \bibinfo {author} {\bibfnamefont {P.}~\bibnamefont {Dai}}, \bibinfo {author} {\bibfnamefont {Z.~Y.}\ \bibnamefont {Meng}}, \bibinfo {author} {\bibfnamefont {Y.-f.}\ \bibnamefont {Yang}}, \bibinfo {author} {\bibfnamefont {H.}~\bibnamefont {Luo}},\ and\ \bibinfo {author} {\bibfnamefont {S.}~\bibnamefont {Li}},\ }\bibfield  {title} {\bibinfo {title} {Effect of nematic order on the low-energy spin fluctuations in detwinned \ce{BaFe_{1.935}Ni_{0.065}As2}},\ }\href {https://doi.org/10.1103/PhysRevLett.117.227003} {\bibfield  {journal} {\bibinfo  {journal} {Phys. Rev. Lett.}\ }\textbf {\bibinfo {volume} {117}},\ \bibinfo {pages} {227003}
  (\bibinfo {year} {2016})}\BibitemShut {NoStop}%
\bibitem [{\citenamefont {Liu}\ \emph {et~al.}(2016)\citenamefont {Liu}, \citenamefont {Gu}, \citenamefont {Zhang}, \citenamefont {Gong}, \citenamefont {Zhang}, \citenamefont {Xie}, \citenamefont {Lu}, \citenamefont {Ma}, \citenamefont {Zhang}, \citenamefont {Zhang}, \citenamefont {Zhu}, \citenamefont {Ren}, \citenamefont {Shan}, \citenamefont {Qiu}, \citenamefont {Dai}, \citenamefont {Yang}, \citenamefont {Luo},\ and\ \citenamefont {Li}}]{157002}%
  \BibitemOpen
  \bibfield  {author} {\bibinfo {author} {\bibfnamefont {Z.}~\bibnamefont {Liu}}, \bibinfo {author} {\bibfnamefont {Y.}~\bibnamefont {Gu}}, \bibinfo {author} {\bibfnamefont {W.}~\bibnamefont {Zhang}}, \bibinfo {author} {\bibfnamefont {D.}~\bibnamefont {Gong}}, \bibinfo {author} {\bibfnamefont {W.}~\bibnamefont {Zhang}}, \bibinfo {author} {\bibfnamefont {T.}~\bibnamefont {Xie}}, \bibinfo {author} {\bibfnamefont {X.}~\bibnamefont {Lu}}, \bibinfo {author} {\bibfnamefont {X.}~\bibnamefont {Ma}}, \bibinfo {author} {\bibfnamefont {X.}~\bibnamefont {Zhang}}, \bibinfo {author} {\bibfnamefont {R.}~\bibnamefont {Zhang}}, \bibinfo {author} {\bibfnamefont {J.}~\bibnamefont {Zhu}}, \bibinfo {author} {\bibfnamefont {C.}~\bibnamefont {Ren}}, \bibinfo {author} {\bibfnamefont {L.}~\bibnamefont {Shan}}, \bibinfo {author} {\bibfnamefont {X.}~\bibnamefont {Qiu}}, \bibinfo {author} {\bibfnamefont {P.}~\bibnamefont {Dai}}, \bibinfo {author} {\bibfnamefont {Y.-f.}\ \bibnamefont {Yang}}, \bibinfo {author} {\bibfnamefont
  {H.}~\bibnamefont {Luo}},\ and\ \bibinfo {author} {\bibfnamefont {S.}~\bibnamefont {Li}},\ }\bibfield  {title} {\bibinfo {title} {Nematic quantum critical fluctuations in \ce{BaFe_{2-x}Ni_{x}As2}},\ }\href {https://doi.org/10.1103/PhysRevLett.117.157002} {\bibfield  {journal} {\bibinfo  {journal} {Phys. Rev. Lett.}\ }\textbf {\bibinfo {volume} {117}},\ \bibinfo {pages} {157002} (\bibinfo {year} {2016})}\BibitemShut {NoStop}%
\bibitem [{\citenamefont {Kuo}\ \emph {et~al.}(2016)\citenamefont {Kuo}, \citenamefont {Chu}, \citenamefont {Palmstrom}, \citenamefont {Kivelson},\ and\ \citenamefont {Fisher}}]{aab0103}%
  \BibitemOpen
  \bibfield  {author} {\bibinfo {author} {\bibfnamefont {H.-H.}\ \bibnamefont {Kuo}}, \bibinfo {author} {\bibfnamefont {J.-H.}\ \bibnamefont {Chu}}, \bibinfo {author} {\bibfnamefont {J.~C.}\ \bibnamefont {Palmstrom}}, \bibinfo {author} {\bibfnamefont {S.~A.}\ \bibnamefont {Kivelson}},\ and\ \bibinfo {author} {\bibfnamefont {I.~R.}\ \bibnamefont {Fisher}},\ }\bibfield  {title} {\bibinfo {title} {Ubiquitous signatures of nematic quantum criticality in optimally doped \ce{Fe}-based superconductors},\ }\href {https://doi.org/10.1126/science.aab0103} {\bibfield  {journal} {\bibinfo  {journal} {Science}\ }\textbf {\bibinfo {volume} {352}},\ \bibinfo {pages} {958} (\bibinfo {year} {2016})}\BibitemShut {NoStop}%
\bibitem [{\citenamefont {Asaba}\ \emph {et~al.}(2024)\citenamefont {Asaba}, \citenamefont {Onishi}, \citenamefont {Kageyama}, \citenamefont {Kiyosue}, \citenamefont {Ohtsuka}, \citenamefont {Suetsugu}, \citenamefont {Kohsaka}, \citenamefont {Gaggl}, \citenamefont {Kasahara}, \citenamefont {Murayama}, \citenamefont {Hashimoto}, \citenamefont {Tazai}, \citenamefont {Kontani}, \citenamefont {Ortiz}, \citenamefont {Wilson}, \citenamefont {Li}, \citenamefont {Wen}, \citenamefont {Shibauchi},\ and\ \citenamefont {Matsuda}}]{Asaba2024}%
  \BibitemOpen
  \bibfield  {author} {\bibinfo {author} {\bibfnamefont {T.}~\bibnamefont {Asaba}}, \bibinfo {author} {\bibfnamefont {A.}~\bibnamefont {Onishi}}, \bibinfo {author} {\bibfnamefont {Y.}~\bibnamefont {Kageyama}}, \bibinfo {author} {\bibfnamefont {T.}~\bibnamefont {Kiyosue}}, \bibinfo {author} {\bibfnamefont {K.}~\bibnamefont {Ohtsuka}}, \bibinfo {author} {\bibfnamefont {S.}~\bibnamefont {Suetsugu}}, \bibinfo {author} {\bibfnamefont {Y.}~\bibnamefont {Kohsaka}}, \bibinfo {author} {\bibfnamefont {T.}~\bibnamefont {Gaggl}}, \bibinfo {author} {\bibfnamefont {Y.}~\bibnamefont {Kasahara}}, \bibinfo {author} {\bibfnamefont {H.}~\bibnamefont {Murayama}}, \bibinfo {author} {\bibfnamefont {K.}~\bibnamefont {Hashimoto}}, \bibinfo {author} {\bibfnamefont {R.}~\bibnamefont {Tazai}}, \bibinfo {author} {\bibfnamefont {H.}~\bibnamefont {Kontani}}, \bibinfo {author} {\bibfnamefont {B.~R.}\ \bibnamefont {Ortiz}}, \bibinfo {author} {\bibfnamefont {S.~D.}\ \bibnamefont {Wilson}}, \bibinfo {author} {\bibfnamefont {Q.}~\bibnamefont
  {Li}}, \bibinfo {author} {\bibfnamefont {H.-H.}\ \bibnamefont {Wen}}, \bibinfo {author} {\bibfnamefont {T.}~\bibnamefont {Shibauchi}},\ and\ \bibinfo {author} {\bibfnamefont {Y.}~\bibnamefont {Matsuda}},\ }\bibfield  {title} {\bibinfo {title} {Evidence for an odd-parity nematic phase above the charge-density-wave transition in a kagome metal},\ }\href {https://doi.org/10.1038/s41567-023-02272-4} {\bibfield  {journal} {\bibinfo  {journal} {Nature Physics}\ }\textbf {\bibinfo {volume} {20}},\ \bibinfo {pages} {40} (\bibinfo {year} {2024})}\BibitemShut {NoStop}%
\bibitem [{\citenamefont {Yang}\ \emph {et~al.}(2024)\citenamefont {Yang}, \citenamefont {Ye}, \citenamefont {Zhao}, \citenamefont {Liu}, \citenamefont {Yi}, \citenamefont {Zhang}, \citenamefont {Xiao}, \citenamefont {Shi}, \citenamefont {You}, \citenamefont {Huang}, \citenamefont {Wang}, \citenamefont {Wang}, \citenamefont {Guo}, \citenamefont {Lin}, \citenamefont {Shen}, \citenamefont {Zhou}, \citenamefont {Chen}, \citenamefont {Dong}, \citenamefont {Su}, \citenamefont {Wang},\ and\ \citenamefont {Gao}}]{Yang2024}%
  \BibitemOpen
  \bibfield  {author} {\bibinfo {author} {\bibfnamefont {H.}~\bibnamefont {Yang}}, \bibinfo {author} {\bibfnamefont {Y.}~\bibnamefont {Ye}}, \bibinfo {author} {\bibfnamefont {Z.}~\bibnamefont {Zhao}}, \bibinfo {author} {\bibfnamefont {J.}~\bibnamefont {Liu}}, \bibinfo {author} {\bibfnamefont {X.-W.}\ \bibnamefont {Yi}}, \bibinfo {author} {\bibfnamefont {Y.}~\bibnamefont {Zhang}}, \bibinfo {author} {\bibfnamefont {H.}~\bibnamefont {Xiao}}, \bibinfo {author} {\bibfnamefont {J.}~\bibnamefont {Shi}}, \bibinfo {author} {\bibfnamefont {J.-Y.}\ \bibnamefont {You}}, \bibinfo {author} {\bibfnamefont {Z.}~\bibnamefont {Huang}}, \bibinfo {author} {\bibfnamefont {B.}~\bibnamefont {Wang}}, \bibinfo {author} {\bibfnamefont {J.}~\bibnamefont {Wang}}, \bibinfo {author} {\bibfnamefont {H.}~\bibnamefont {Guo}}, \bibinfo {author} {\bibfnamefont {X.}~\bibnamefont {Lin}}, \bibinfo {author} {\bibfnamefont {C.}~\bibnamefont {Shen}}, \bibinfo {author} {\bibfnamefont {W.}~\bibnamefont {Zhou}}, \bibinfo {author} {\bibfnamefont
  {H.}~\bibnamefont {Chen}}, \bibinfo {author} {\bibfnamefont {X.}~\bibnamefont {Dong}}, \bibinfo {author} {\bibfnamefont {G.}~\bibnamefont {Su}}, \bibinfo {author} {\bibfnamefont {Z.}~\bibnamefont {Wang}},\ and\ \bibinfo {author} {\bibfnamefont {H.-J.}\ \bibnamefont {Gao}},\ }\bibfield  {title} {\bibinfo {title} {Superconductivity and nematic order in a new titanium-based kagome metal \ce{CsTi3Bi5} without charge density wave order},\ }\href {https://doi.org/10.1038/s41467-024-53870-6} {\bibfield  {journal} {\bibinfo  {journal} {Nature Communications}\ }\textbf {\bibinfo {volume} {15}},\ \bibinfo {pages} {9626} (\bibinfo {year} {2024})}\BibitemShut {NoStop}%
\bibitem [{\citenamefont {Jiang}\ \emph {et~al.}(2023)\citenamefont {Jiang}, \citenamefont {Ma}, \citenamefont {Xia}, \citenamefont {Liu}, \citenamefont {Xiao}, \citenamefont {Liu}, \citenamefont {Yang}, \citenamefont {Ding}, \citenamefont {Huang}, \citenamefont {Liu}, \citenamefont {Qiao}, \citenamefont {Liu}, \citenamefont {Peng}, \citenamefont {Cho}, \citenamefont {Guo}, \citenamefont {Liu},\ and\ \citenamefont {Shen}}]{Jiang2023}%
  \BibitemOpen
  \bibfield  {author} {\bibinfo {author} {\bibfnamefont {Z.}~\bibnamefont {Jiang}}, \bibinfo {author} {\bibfnamefont {H.}~\bibnamefont {Ma}}, \bibinfo {author} {\bibfnamefont {W.}~\bibnamefont {Xia}}, \bibinfo {author} {\bibfnamefont {Z.}~\bibnamefont {Liu}}, \bibinfo {author} {\bibfnamefont {Q.}~\bibnamefont {Xiao}}, \bibinfo {author} {\bibfnamefont {Z.}~\bibnamefont {Liu}}, \bibinfo {author} {\bibfnamefont {Y.}~\bibnamefont {Yang}}, \bibinfo {author} {\bibfnamefont {J.}~\bibnamefont {Ding}}, \bibinfo {author} {\bibfnamefont {Z.}~\bibnamefont {Huang}}, \bibinfo {author} {\bibfnamefont {J.}~\bibnamefont {Liu}}, \bibinfo {author} {\bibfnamefont {Y.}~\bibnamefont {Qiao}}, \bibinfo {author} {\bibfnamefont {J.}~\bibnamefont {Liu}}, \bibinfo {author} {\bibfnamefont {Y.}~\bibnamefont {Peng}}, \bibinfo {author} {\bibfnamefont {S.}~\bibnamefont {Cho}}, \bibinfo {author} {\bibfnamefont {Y.}~\bibnamefont {Guo}}, \bibinfo {author} {\bibfnamefont {J.}~\bibnamefont {Liu}},\ and\ \bibinfo {author} {\bibfnamefont
  {D.}~\bibnamefont {Shen}},\ }\bibfield  {title} {\bibinfo {title} {Observation of electronic nematicity driven by the three-dimensional charge density wave in kagome lattice \ce{KV3Sb5}},\ }\href {https://doi.org/10.1021/acs.nanolett.3c01151} {\bibfield  {journal} {\bibinfo  {journal} {Nano Letters}\ }\textbf {\bibinfo {volume} {23}},\ \bibinfo {pages} {5625} (\bibinfo {year} {2023})}\BibitemShut {NoStop}%
\bibitem [{\citenamefont {Cheng}\ \emph {et~al.}(2025)\citenamefont {Cheng}, \citenamefont {Zhang}, \citenamefont {Sun}, \citenamefont {Chen}, \citenamefont {Qin}, \citenamefont {Ren}, \citenamefont {Cao}, \citenamefont {Xie},\ and\ \citenamefont {Wu}}]{021018}%
  \BibitemOpen
  \bibfield  {author} {\bibinfo {author} {\bibfnamefont {X.~B.}\ \bibnamefont {Cheng}}, \bibinfo {author} {\bibfnamefont {M.}~\bibnamefont {Zhang}}, \bibinfo {author} {\bibfnamefont {Y.~Q.}\ \bibnamefont {Sun}}, \bibinfo {author} {\bibfnamefont {G.~F.}\ \bibnamefont {Chen}}, \bibinfo {author} {\bibfnamefont {M.}~\bibnamefont {Qin}}, \bibinfo {author} {\bibfnamefont {T.~S.}\ \bibnamefont {Ren}}, \bibinfo {author} {\bibfnamefont {X.~S.}\ \bibnamefont {Cao}}, \bibinfo {author} {\bibfnamefont {Y.~W.}\ \bibnamefont {Xie}},\ and\ \bibinfo {author} {\bibfnamefont {J.}~\bibnamefont {Wu}},\ }\bibfield  {title} {\bibinfo {title} {Electronic nematicity in interface superconducting $\mathrm{LAO}/\mathrm{KTO}(111)$},\ }\href {https://doi.org/10.1103/PhysRevX.15.021018} {\bibfield  {journal} {\bibinfo  {journal} {Phys. Rev. X}\ }\textbf {\bibinfo {volume} {15}},\ \bibinfo {pages} {021018} (\bibinfo {year} {2025})}\BibitemShut {NoStop}%
\bibitem [{\citenamefont {Naritsuka}\ \emph {et~al.}(2023)\citenamefont {Naritsuka}, \citenamefont {Benedičič}, \citenamefont {Rhodes}, \citenamefont {Marques}, \citenamefont {Trainer}, \citenamefont {Li}, \citenamefont {Komarek},\ and\ \citenamefont {Wahl}}]{2308972120}%
  \BibitemOpen
  \bibfield  {author} {\bibinfo {author} {\bibfnamefont {M.}~\bibnamefont {Naritsuka}}, \bibinfo {author} {\bibfnamefont {I.}~\bibnamefont {Benedičič}}, \bibinfo {author} {\bibfnamefont {L.~C.}\ \bibnamefont {Rhodes}}, \bibinfo {author} {\bibfnamefont {C.~A.}\ \bibnamefont {Marques}}, \bibinfo {author} {\bibfnamefont {C.}~\bibnamefont {Trainer}}, \bibinfo {author} {\bibfnamefont {Z.}~\bibnamefont {Li}}, \bibinfo {author} {\bibfnamefont {A.~C.}\ \bibnamefont {Komarek}},\ and\ \bibinfo {author} {\bibfnamefont {P.}~\bibnamefont {Wahl}},\ }\bibfield  {title} {\bibinfo {title} {Compass-like manipulation of electronic nematicity in \ce{Sr3Ru2O7}},\ }\href {https://doi.org/https://doi.org/10.1073/pnas.2308972120} {\bibfield  {journal} {\bibinfo  {journal} {Proceedings of the National Academy of Sciences}\ }\textbf {\bibinfo {volume} {120}},\ \bibinfo {pages} {e2308972120} (\bibinfo {year} {2023})}\BibitemShut {NoStop}%
\bibitem [{\citenamefont {Du}\ \emph {et~al.}(2021)\citenamefont {Du}, \citenamefont {Hasan}, \citenamefont {Castellanos-Gomez}, \citenamefont {Liu}, \citenamefont {Yao}, \citenamefont {Lau},\ and\ \citenamefont {Sun}}]{Du2021}%
  \BibitemOpen
  \bibfield  {author} {\bibinfo {author} {\bibfnamefont {L.}~\bibnamefont {Du}}, \bibinfo {author} {\bibfnamefont {T.}~\bibnamefont {Hasan}}, \bibinfo {author} {\bibfnamefont {A.}~\bibnamefont {Castellanos-Gomez}}, \bibinfo {author} {\bibfnamefont {G.-B.}\ \bibnamefont {Liu}}, \bibinfo {author} {\bibfnamefont {Y.}~\bibnamefont {Yao}}, \bibinfo {author} {\bibfnamefont {C.~N.}\ \bibnamefont {Lau}},\ and\ \bibinfo {author} {\bibfnamefont {Z.}~\bibnamefont {Sun}},\ }\bibfield  {title} {\bibinfo {title} {Engineering symmetry breaking in \ce{2D} layered materials},\ }\href {https://doi.org/10.1038/s42254-020-00276-0} {\bibfield  {journal} {\bibinfo  {journal} {Nature Reviews Physics}\ }\textbf {\bibinfo {volume} {3}},\ \bibinfo {pages} {193} (\bibinfo {year} {2021})}\BibitemShut {NoStop}%
\bibitem [{\citenamefont {Damascelli}\ \emph {et~al.}(2000)\citenamefont {Damascelli}, \citenamefont {Lu}, \citenamefont {Shen}, \citenamefont {Armitage}, \citenamefont {Ronning}, \citenamefont {Feng}, \citenamefont {Kim}, \citenamefont {Shen}, \citenamefont {Kimura}, \citenamefont {Tokura}, \citenamefont {Mao},\ and\ \citenamefont {Maeno}}]{5194}%
  \BibitemOpen
  \bibfield  {author} {\bibinfo {author} {\bibfnamefont {A.}~\bibnamefont {Damascelli}}, \bibinfo {author} {\bibfnamefont {D.~H.}\ \bibnamefont {Lu}}, \bibinfo {author} {\bibfnamefont {K.~M.}\ \bibnamefont {Shen}}, \bibinfo {author} {\bibfnamefont {N.~P.}\ \bibnamefont {Armitage}}, \bibinfo {author} {\bibfnamefont {F.}~\bibnamefont {Ronning}}, \bibinfo {author} {\bibfnamefont {D.~L.}\ \bibnamefont {Feng}}, \bibinfo {author} {\bibfnamefont {C.}~\bibnamefont {Kim}}, \bibinfo {author} {\bibfnamefont {Z.-X.}\ \bibnamefont {Shen}}, \bibinfo {author} {\bibfnamefont {T.}~\bibnamefont {Kimura}}, \bibinfo {author} {\bibfnamefont {Y.}~\bibnamefont {Tokura}}, \bibinfo {author} {\bibfnamefont {Z.~Q.}\ \bibnamefont {Mao}},\ and\ \bibinfo {author} {\bibfnamefont {Y.}~\bibnamefont {Maeno}},\ }\bibfield  {title} {\bibinfo {title} {Fermi surface, surface states, and surface reconstruction in \ce{Sr2RuO4}},\ }\href {https://doi.org/10.1103/PhysRevLett.85.5194} {\bibfield  {journal} {\bibinfo  {journal} {Phys. Rev. Lett.}\
  }\textbf {\bibinfo {volume} {85}},\ \bibinfo {pages} {5194} (\bibinfo {year} {2000})}\BibitemShut {NoStop}%
\bibitem [{\citenamefont {Wang}\ \emph {et~al.}(2022)\citenamefont {Wang}, \citenamefont {Gao}, \citenamefont {Yang}, \citenamefont {Gu}, \citenamefont {Lu}, \citenamefont {Zhang}, \citenamefont {Gao}, \citenamefont {Ren}, \citenamefont {Dong}, \citenamefont {Jiang}, \citenamefont {Watanabe}, \citenamefont {Taniguchi}, \citenamefont {Kang}, \citenamefont {Lou}, \citenamefont {Mao}, \citenamefont {Liu}, \citenamefont {Ye}, \citenamefont {Han}, \citenamefont {Chang}, \citenamefont {Zhang},\ and\ \citenamefont {Zhang}}]{Wang2022222}%
  \BibitemOpen
  \bibfield  {author} {\bibinfo {author} {\bibfnamefont {Y.}~\bibnamefont {Wang}}, \bibinfo {author} {\bibfnamefont {X.}~\bibnamefont {Gao}}, \bibinfo {author} {\bibfnamefont {K.}~\bibnamefont {Yang}}, \bibinfo {author} {\bibfnamefont {P.}~\bibnamefont {Gu}}, \bibinfo {author} {\bibfnamefont {X.}~\bibnamefont {Lu}}, \bibinfo {author} {\bibfnamefont {S.}~\bibnamefont {Zhang}}, \bibinfo {author} {\bibfnamefont {Y.}~\bibnamefont {Gao}}, \bibinfo {author} {\bibfnamefont {N.}~\bibnamefont {Ren}}, \bibinfo {author} {\bibfnamefont {B.}~\bibnamefont {Dong}}, \bibinfo {author} {\bibfnamefont {Y.}~\bibnamefont {Jiang}}, \bibinfo {author} {\bibfnamefont {K.}~\bibnamefont {Watanabe}}, \bibinfo {author} {\bibfnamefont {T.}~\bibnamefont {Taniguchi}}, \bibinfo {author} {\bibfnamefont {J.}~\bibnamefont {Kang}}, \bibinfo {author} {\bibfnamefont {W.}~\bibnamefont {Lou}}, \bibinfo {author} {\bibfnamefont {J.}~\bibnamefont {Mao}}, \bibinfo {author} {\bibfnamefont {J.}~\bibnamefont {Liu}}, \bibinfo {author} {\bibfnamefont
  {Y.}~\bibnamefont {Ye}}, \bibinfo {author} {\bibfnamefont {Z.}~\bibnamefont {Han}}, \bibinfo {author} {\bibfnamefont {K.}~\bibnamefont {Chang}}, \bibinfo {author} {\bibfnamefont {J.}~\bibnamefont {Zhang}},\ and\ \bibinfo {author} {\bibfnamefont {Z.}~\bibnamefont {Zhang}},\ }\bibfield  {title} {\bibinfo {title} {Quantum hall phase in graphene engineered by interfacial charge coupling},\ }\href {https://doi.org/10.1038/s41565-022-01248-4} {\bibfield  {journal} {\bibinfo  {journal} {Nature Nanotechnology}\ }\textbf {\bibinfo {volume} {17}},\ \bibinfo {pages} {1272} (\bibinfo {year} {2022})}\BibitemShut {NoStop}%
\bibitem [{\citenamefont {Yuan}\ \emph {et~al.}(2019)\citenamefont {Yuan}, \citenamefont {Pan}, \citenamefont {Wang}, \citenamefont {Fang}, \citenamefont {Song}, \citenamefont {Wang}, \citenamefont {He}, \citenamefont {Ma}, \citenamefont {Zhang}, \citenamefont {Huang}, \citenamefont {Li},\ and\ \citenamefont {Xue}}]{Yuan2019}%
  \BibitemOpen
  \bibfield  {author} {\bibinfo {author} {\bibfnamefont {Y.}~\bibnamefont {Yuan}}, \bibinfo {author} {\bibfnamefont {J.}~\bibnamefont {Pan}}, \bibinfo {author} {\bibfnamefont {X.}~\bibnamefont {Wang}}, \bibinfo {author} {\bibfnamefont {Y.}~\bibnamefont {Fang}}, \bibinfo {author} {\bibfnamefont {C.}~\bibnamefont {Song}}, \bibinfo {author} {\bibfnamefont {L.}~\bibnamefont {Wang}}, \bibinfo {author} {\bibfnamefont {K.}~\bibnamefont {He}}, \bibinfo {author} {\bibfnamefont {X.}~\bibnamefont {Ma}}, \bibinfo {author} {\bibfnamefont {H.}~\bibnamefont {Zhang}}, \bibinfo {author} {\bibfnamefont {F.}~\bibnamefont {Huang}}, \bibinfo {author} {\bibfnamefont {W.}~\bibnamefont {Li}},\ and\ \bibinfo {author} {\bibfnamefont {Q.-K.}\ \bibnamefont {Xue}},\ }\bibfield  {title} {\bibinfo {title} {Evidence of anisotropic majorana bound states in 2\ce{M}-\ce{WS2}},\ }\href {https://doi.org/10.1038/s41567-019-0576-7} {\bibfield  {journal} {\bibinfo  {journal} {Nature Physics}\ }\textbf {\bibinfo {volume} {15}},\ \bibinfo {pages}
  {1046} (\bibinfo {year} {2019})}\BibitemShut {NoStop}%
\bibitem [{\citenamefont {Chuang}\ \emph {et~al.}(2010)\citenamefont {Chuang}, \citenamefont {Allan}, \citenamefont {Lee}, \citenamefont {Xie}, \citenamefont {Ni}, \citenamefont {Bud’ko}, \citenamefont {Boebinger}, \citenamefont {Canfield},\ and\ \citenamefont {Davis}}]{1181083}%
  \BibitemOpen
  \bibfield  {author} {\bibinfo {author} {\bibfnamefont {T.-M.}\ \bibnamefont {Chuang}}, \bibinfo {author} {\bibfnamefont {M.~P.}\ \bibnamefont {Allan}}, \bibinfo {author} {\bibfnamefont {J.}~\bibnamefont {Lee}}, \bibinfo {author} {\bibfnamefont {Y.}~\bibnamefont {Xie}}, \bibinfo {author} {\bibfnamefont {N.}~\bibnamefont {Ni}}, \bibinfo {author} {\bibfnamefont {S.~L.}\ \bibnamefont {Bud’ko}}, \bibinfo {author} {\bibfnamefont {G.~S.}\ \bibnamefont {Boebinger}}, \bibinfo {author} {\bibfnamefont {P.~C.}\ \bibnamefont {Canfield}},\ and\ \bibinfo {author} {\bibfnamefont {J.~C.}\ \bibnamefont {Davis}},\ }\bibfield  {title} {\bibinfo {title} {Nematic electronic structure in the "parent" state of the iron-based superconductor \ce{Ca(Fe_{1-x}Co_{x})2As_{2}}},\ }\href {https://doi.org/10.1126/science.1181083} {\bibfield  {journal} {\bibinfo  {journal} {Science}\ }\textbf {\bibinfo {volume} {327}},\ \bibinfo {pages} {181} (\bibinfo {year} {2010})}\BibitemShut {NoStop}%
\bibitem [{\citenamefont {Shang}\ \emph {et~al.}(2021)\citenamefont {Shang}, \citenamefont {Xu}, \citenamefont {Gawryluk}, \citenamefont {Ma}, \citenamefont {Shiroka}, \citenamefont {Shi},\ and\ \citenamefont {Pomjakushina}}]{L020405}%
  \BibitemOpen
  \bibfield  {author} {\bibinfo {author} {\bibfnamefont {T.}~\bibnamefont {Shang}}, \bibinfo {author} {\bibfnamefont {Y.}~\bibnamefont {Xu}}, \bibinfo {author} {\bibfnamefont {D.~J.}\ \bibnamefont {Gawryluk}}, \bibinfo {author} {\bibfnamefont {J.~Z.}\ \bibnamefont {Ma}}, \bibinfo {author} {\bibfnamefont {T.}~\bibnamefont {Shiroka}}, \bibinfo {author} {\bibfnamefont {M.}~\bibnamefont {Shi}},\ and\ \bibinfo {author} {\bibfnamefont {E.}~\bibnamefont {Pomjakushina}},\ }\bibfield  {title} {\bibinfo {title} {Anomalous hall resistivity and possible topological hall effect in the \ce{EuAl4} antiferromagnet},\ }\href {https://doi.org/10.1103/PhysRevB.103.L020405} {\bibfield  {journal} {\bibinfo  {journal} {Phys. Rev. B}\ }\textbf {\bibinfo {volume} {103}},\ \bibinfo {pages} {L020405} (\bibinfo {year} {2021})}\BibitemShut {NoStop}%
\bibitem [{\citenamefont {Moya}\ \emph {et~al.}(2023)\citenamefont {Moya}, \citenamefont {Huang}, \citenamefont {Lei}, \citenamefont {Allen}, \citenamefont {Gao}, \citenamefont {Sun}, \citenamefont {Yi},\ and\ \citenamefont {Morosan}}]{064436}%
  \BibitemOpen
  \bibfield  {author} {\bibinfo {author} {\bibfnamefont {J.~M.}\ \bibnamefont {Moya}}, \bibinfo {author} {\bibfnamefont {J.}~\bibnamefont {Huang}}, \bibinfo {author} {\bibfnamefont {S.}~\bibnamefont {Lei}}, \bibinfo {author} {\bibfnamefont {K.}~\bibnamefont {Allen}}, \bibinfo {author} {\bibfnamefont {Y.}~\bibnamefont {Gao}}, \bibinfo {author} {\bibfnamefont {Y.}~\bibnamefont {Sun}}, \bibinfo {author} {\bibfnamefont {M.}~\bibnamefont {Yi}},\ and\ \bibinfo {author} {\bibfnamefont {E.}~\bibnamefont {Morosan}},\ }\bibfield  {title} {\bibinfo {title} {Real-space and reciprocal-space topology in the ${{\mathrm{Eu}(\mathrm{Ga}}_{1\ensuremath{-}x}{\mathrm{Al}}_{x})}_{4}$ square net system},\ }\href {https://doi.org/10.1103/PhysRevB.108.064436} {\bibfield  {journal} {\bibinfo  {journal} {Phys. Rev. B}\ }\textbf {\bibinfo {volume} {108}},\ \bibinfo {pages} {064436} (\bibinfo {year} {2023})}\BibitemShut {NoStop}%
\bibitem [{\citenamefont {Lei}\ \emph {et~al.}(2023)\citenamefont {Lei}, \citenamefont {Allen}, \citenamefont {Huang}, \citenamefont {Moya}, \citenamefont {Wu}, \citenamefont {Casas}, \citenamefont {Zhang}, \citenamefont {Oh}, \citenamefont {Hashimoto}, \citenamefont {Lu}, \citenamefont {Denlinger}, \citenamefont {Jozwiak}, \citenamefont {Bostwick}, \citenamefont {Rotenberg}, \citenamefont {Balicas}, \citenamefont {Birgeneau}, \citenamefont {Foster}, \citenamefont {Yi}, \citenamefont {Sun},\ and\ \citenamefont {Morosan}}]{Lei2023}%
  \BibitemOpen
  \bibfield  {author} {\bibinfo {author} {\bibfnamefont {S.}~\bibnamefont {Lei}}, \bibinfo {author} {\bibfnamefont {K.}~\bibnamefont {Allen}}, \bibinfo {author} {\bibfnamefont {J.}~\bibnamefont {Huang}}, \bibinfo {author} {\bibfnamefont {J.~M.}\ \bibnamefont {Moya}}, \bibinfo {author} {\bibfnamefont {T.~C.}\ \bibnamefont {Wu}}, \bibinfo {author} {\bibfnamefont {B.}~\bibnamefont {Casas}}, \bibinfo {author} {\bibfnamefont {Y.}~\bibnamefont {Zhang}}, \bibinfo {author} {\bibfnamefont {J.~S.}\ \bibnamefont {Oh}}, \bibinfo {author} {\bibfnamefont {M.}~\bibnamefont {Hashimoto}}, \bibinfo {author} {\bibfnamefont {D.}~\bibnamefont {Lu}}, \bibinfo {author} {\bibfnamefont {J.}~\bibnamefont {Denlinger}}, \bibinfo {author} {\bibfnamefont {C.}~\bibnamefont {Jozwiak}}, \bibinfo {author} {\bibfnamefont {A.}~\bibnamefont {Bostwick}}, \bibinfo {author} {\bibfnamefont {E.}~\bibnamefont {Rotenberg}}, \bibinfo {author} {\bibfnamefont {L.}~\bibnamefont {Balicas}}, \bibinfo {author} {\bibfnamefont {R.}~\bibnamefont {Birgeneau}},
  \bibinfo {author} {\bibfnamefont {M.~S.}\ \bibnamefont {Foster}}, \bibinfo {author} {\bibfnamefont {M.}~\bibnamefont {Yi}}, \bibinfo {author} {\bibfnamefont {Y.}~\bibnamefont {Sun}},\ and\ \bibinfo {author} {\bibfnamefont {E.}~\bibnamefont {Morosan}},\ }\bibfield  {title} {\bibinfo {title} {Weyl nodal ring states and landau quantization with very large magnetoresistance in square-net magnet \ce{EuGa4}},\ }\href {https://doi.org/10.1038/s41467-023-40767-z} {\bibfield  {journal} {\bibinfo  {journal} {Nature Communications}\ }\textbf {\bibinfo {volume} {14}},\ \bibinfo {pages} {5812} (\bibinfo {year} {2023})}\BibitemShut {NoStop}%
\bibitem [{\citenamefont {Wang}\ \emph {et~al.}(2021)\citenamefont {Wang}, \citenamefont {Mori}, \citenamefont {Wang}, \citenamefont {Wang}, \citenamefont {Ma}, \citenamefont {Latzke}, \citenamefont {Graf}, \citenamefont {Denlinger}, \citenamefont {Campbell}, \citenamefont {Bernevig}, \citenamefont {Lanzara},\ and\ \citenamefont {Paglione}}]{Wang2021}%
  \BibitemOpen
  \bibfield  {author} {\bibinfo {author} {\bibfnamefont {K.}~\bibnamefont {Wang}}, \bibinfo {author} {\bibfnamefont {R.}~\bibnamefont {Mori}}, \bibinfo {author} {\bibfnamefont {Z.}~\bibnamefont {Wang}}, \bibinfo {author} {\bibfnamefont {L.}~\bibnamefont {Wang}}, \bibinfo {author} {\bibfnamefont {J.~H.~S.}\ \bibnamefont {Ma}}, \bibinfo {author} {\bibfnamefont {D.~W.}\ \bibnamefont {Latzke}}, \bibinfo {author} {\bibfnamefont {D.~E.}\ \bibnamefont {Graf}}, \bibinfo {author} {\bibfnamefont {J.~D.}\ \bibnamefont {Denlinger}}, \bibinfo {author} {\bibfnamefont {D.}~\bibnamefont {Campbell}}, \bibinfo {author} {\bibfnamefont {B.~A.}\ \bibnamefont {Bernevig}}, \bibinfo {author} {\bibfnamefont {A.}~\bibnamefont {Lanzara}},\ and\ \bibinfo {author} {\bibfnamefont {J.}~\bibnamefont {Paglione}},\ }\bibfield  {title} {\bibinfo {title} {Crystalline symmetry-protected non-trivial topology in prototype compound \ce{BaAl4}},\ }\href {https://doi.org/10.1038/s41535-021-00325-6} {\bibfield  {journal} {\bibinfo  {journal} {npj
  Quantum Materials}\ }\textbf {\bibinfo {volume} {6}},\ \bibinfo {pages} {28} (\bibinfo {year} {2021})}\BibitemShut {NoStop}%
\bibitem [{\citenamefont {Takagi}\ \emph {et~al.}(2022)\citenamefont {Takagi}, \citenamefont {Matsuyama}, \citenamefont {Ukleev}, \citenamefont {Yu}, \citenamefont {White}, \citenamefont {Francoual}, \citenamefont {Mardegan}, \citenamefont {Hayami}, \citenamefont {Saito}, \citenamefont {Kaneko}, \citenamefont {Ohishi}, \citenamefont {{\={O}}nuki}, \citenamefont {Arima}, \citenamefont {Tokura}, \citenamefont {Nakajima},\ and\ \citenamefont {Seki}}]{Takagi2022}%
  \BibitemOpen
  \bibfield  {author} {\bibinfo {author} {\bibfnamefont {R.}~\bibnamefont {Takagi}}, \bibinfo {author} {\bibfnamefont {N.}~\bibnamefont {Matsuyama}}, \bibinfo {author} {\bibfnamefont {V.}~\bibnamefont {Ukleev}}, \bibinfo {author} {\bibfnamefont {L.}~\bibnamefont {Yu}}, \bibinfo {author} {\bibfnamefont {J.~S.}\ \bibnamefont {White}}, \bibinfo {author} {\bibfnamefont {S.}~\bibnamefont {Francoual}}, \bibinfo {author} {\bibfnamefont {J.~R.~L.}\ \bibnamefont {Mardegan}}, \bibinfo {author} {\bibfnamefont {S.}~\bibnamefont {Hayami}}, \bibinfo {author} {\bibfnamefont {H.}~\bibnamefont {Saito}}, \bibinfo {author} {\bibfnamefont {K.}~\bibnamefont {Kaneko}}, \bibinfo {author} {\bibfnamefont {K.}~\bibnamefont {Ohishi}}, \bibinfo {author} {\bibfnamefont {Y.}~\bibnamefont {{\={O}}nuki}}, \bibinfo {author} {\bibfnamefont {T.-h.}\ \bibnamefont {Arima}}, \bibinfo {author} {\bibfnamefont {Y.}~\bibnamefont {Tokura}}, \bibinfo {author} {\bibfnamefont {T.}~\bibnamefont {Nakajima}},\ and\ \bibinfo {author} {\bibfnamefont
  {S.}~\bibnamefont {Seki}},\ }\bibfield  {title} {\bibinfo {title} {Square and rhombic lattices of magnetic skyrmions in a centrosymmetric binary compound},\ }\href {https://doi.org/10.1038/s41467-022-29131-9} {\bibfield  {journal} {\bibinfo  {journal} {Nature Communications}\ }\textbf {\bibinfo {volume} {13}},\ \bibinfo {pages} {1472} (\bibinfo {year} {2022})}\BibitemShut {NoStop}%
\bibitem [{\citenamefont {Miao}\ \emph {et~al.}(2024)\citenamefont {Miao}, \citenamefont {Bouaziz}, \citenamefont {Fabbris}, \citenamefont {Meier}, \citenamefont {Yang}, \citenamefont {Li}, \citenamefont {Nelson}, \citenamefont {Vescovo}, \citenamefont {Zhang}, \citenamefont {Christianson}, \citenamefont {Lee}, \citenamefont {Zhang}, \citenamefont {Batista},\ and\ \citenamefont {Bl\"ugel}}]{011053}%
  \BibitemOpen
  \bibfield  {author} {\bibinfo {author} {\bibfnamefont {H.}~\bibnamefont {Miao}}, \bibinfo {author} {\bibfnamefont {J.}~\bibnamefont {Bouaziz}}, \bibinfo {author} {\bibfnamefont {G.}~\bibnamefont {Fabbris}}, \bibinfo {author} {\bibfnamefont {W.~R.}\ \bibnamefont {Meier}}, \bibinfo {author} {\bibfnamefont {F.~Z.}\ \bibnamefont {Yang}}, \bibinfo {author} {\bibfnamefont {H.~X.}\ \bibnamefont {Li}}, \bibinfo {author} {\bibfnamefont {C.}~\bibnamefont {Nelson}}, \bibinfo {author} {\bibfnamefont {E.}~\bibnamefont {Vescovo}}, \bibinfo {author} {\bibfnamefont {S.}~\bibnamefont {Zhang}}, \bibinfo {author} {\bibfnamefont {A.~D.}\ \bibnamefont {Christianson}}, \bibinfo {author} {\bibfnamefont {H.~N.}\ \bibnamefont {Lee}}, \bibinfo {author} {\bibfnamefont {Y.}~\bibnamefont {Zhang}}, \bibinfo {author} {\bibfnamefont {C.~D.}\ \bibnamefont {Batista}},\ and\ \bibinfo {author} {\bibfnamefont {S.}~\bibnamefont {Bl\"ugel}},\ }\bibfield  {title} {\bibinfo {title} {Spontaneous chirality flipping in an orthogonal spin-charge
  ordered topological magnet},\ }\href {https://doi.org/10.1103/PhysRevX.14.011053} {\bibfield  {journal} {\bibinfo  {journal} {Phys. Rev. X}\ }\textbf {\bibinfo {volume} {14}},\ \bibinfo {pages} {011053} (\bibinfo {year} {2024})}\BibitemShut {NoStop}%
\bibitem [{\citenamefont {Vibhakar}\ \emph {et~al.}(2024)\citenamefont {Vibhakar}, \citenamefont {Khalyavin}, \citenamefont {Orlandi}, \citenamefont {Moya}, \citenamefont {Lei}, \citenamefont {Morosan},\ and\ \citenamefont {Bombardi}}]{Vibhakar2024}%
  \BibitemOpen
  \bibfield  {author} {\bibinfo {author} {\bibfnamefont {A.~M.}\ \bibnamefont {Vibhakar}}, \bibinfo {author} {\bibfnamefont {D.~D.}\ \bibnamefont {Khalyavin}}, \bibinfo {author} {\bibfnamefont {F.}~\bibnamefont {Orlandi}}, \bibinfo {author} {\bibfnamefont {J.~M.}\ \bibnamefont {Moya}}, \bibinfo {author} {\bibfnamefont {S.}~\bibnamefont {Lei}}, \bibinfo {author} {\bibfnamefont {E.}~\bibnamefont {Morosan}},\ and\ \bibinfo {author} {\bibfnamefont {A.}~\bibnamefont {Bombardi}},\ }\bibfield  {title} {\bibinfo {title} {Spontaneous reversal of spin chirality and competing phases in the topological magnet \ce{EuAl4}},\ }\href {https://doi.org/10.1038/s42005-024-01802-7} {\bibfield  {journal} {\bibinfo  {journal} {Communications Physics}\ }\textbf {\bibinfo {volume} {7}},\ \bibinfo {pages} {313} (\bibinfo {year} {2024})}\BibitemShut {NoStop}%
\bibitem [{\citenamefont {Araki}\ \emph {et~al.}(2014)\citenamefont {Araki}, \citenamefont {Ikeda}, \citenamefont {Kobayashi}, \citenamefont {Nakamura}, \citenamefont {Hiranaka}, \citenamefont {Hedo}, \citenamefont {Nakama},\ and\ \citenamefont {Ōnuki}}]{015001}%
  \BibitemOpen
  \bibfield  {author} {\bibinfo {author} {\bibfnamefont {S.}~\bibnamefont {Araki}}, \bibinfo {author} {\bibfnamefont {Y.}~\bibnamefont {Ikeda}}, \bibinfo {author} {\bibfnamefont {T.~C.}\ \bibnamefont {Kobayashi}}, \bibinfo {author} {\bibfnamefont {A.}~\bibnamefont {Nakamura}}, \bibinfo {author} {\bibfnamefont {Y.}~\bibnamefont {Hiranaka}}, \bibinfo {author} {\bibfnamefont {M.}~\bibnamefont {Hedo}}, \bibinfo {author} {\bibfnamefont {T.}~\bibnamefont {Nakama}},\ and\ \bibinfo {author} {\bibfnamefont {Y.}~\bibnamefont {Ōnuki}},\ }\bibfield  {title} {\bibinfo {title} {Charge density wave transition in \ce{EuAl4}},\ }\href {https://doi.org/https://doi.org/10.7566/JPSJ.83.015001} {\bibfield  {journal} {\bibinfo  {journal} {Journal of the Physical Society of Japan}\ }\textbf {\bibinfo {volume} {83}},\ \bibinfo {pages} {015001} (\bibinfo {year} {2014})}\BibitemShut {NoStop}%
\bibitem [{\citenamefont {Nakamura}\ \emph {et~al.}(2015)\citenamefont {Nakamura}, \citenamefont {Uejo}, \citenamefont {Honda}, \citenamefont {Takeuchi}, \citenamefont {Harima}, \citenamefont {Yamamoto}, \citenamefont {Haga}, \citenamefont {Matsubayashi}, \citenamefont {Uwatoko}, \citenamefont {Hedo}, \citenamefont {Nakama},\ and\ \citenamefont {Ōnuki}}]{124711}%
  \BibitemOpen
  \bibfield  {author} {\bibinfo {author} {\bibfnamefont {A.}~\bibnamefont {Nakamura}}, \bibinfo {author} {\bibfnamefont {T.}~\bibnamefont {Uejo}}, \bibinfo {author} {\bibfnamefont {F.}~\bibnamefont {Honda}}, \bibinfo {author} {\bibfnamefont {T.}~\bibnamefont {Takeuchi}}, \bibinfo {author} {\bibfnamefont {H.}~\bibnamefont {Harima}}, \bibinfo {author} {\bibfnamefont {E.}~\bibnamefont {Yamamoto}}, \bibinfo {author} {\bibfnamefont {Y.}~\bibnamefont {Haga}}, \bibinfo {author} {\bibfnamefont {K.}~\bibnamefont {Matsubayashi}}, \bibinfo {author} {\bibfnamefont {Y.}~\bibnamefont {Uwatoko}}, \bibinfo {author} {\bibfnamefont {M.}~\bibnamefont {Hedo}}, \bibinfo {author} {\bibfnamefont {T.}~\bibnamefont {Nakama}},\ and\ \bibinfo {author} {\bibfnamefont {Y.}~\bibnamefont {Ōnuki}},\ }\bibfield  {title} {\bibinfo {title} {Transport and magnetic properties of \ce{EuAl4} and \ce{EuGa4}},\ }\href {https://doi.org/https://doi.org/10.7566/JPSJ.84.124711} {\bibfield  {journal} {\bibinfo  {journal} {Journal of the Physical Society
  of Japan}\ }\textbf {\bibinfo {volume} {84}},\ \bibinfo {pages} {124711} (\bibinfo {year} {2015})}\BibitemShut {NoStop}%
\bibitem [{\citenamefont {Kobata}\ \emph {et~al.}(2016)\citenamefont {Kobata}, \citenamefont {Fujimori}, \citenamefont {Takeda}, \citenamefont {Okane}, \citenamefont {Saitoh}, \citenamefont {Kobayashi}, \citenamefont {Yamagami}, \citenamefont {Nakamura}, \citenamefont {Hedo}, \citenamefont {Nakama},\ and\ \citenamefont {Ōnuki}}]{094703}%
  \BibitemOpen
  \bibfield  {author} {\bibinfo {author} {\bibfnamefont {M.}~\bibnamefont {Kobata}}, \bibinfo {author} {\bibfnamefont {S.-i.}\ \bibnamefont {Fujimori}}, \bibinfo {author} {\bibfnamefont {Y.}~\bibnamefont {Takeda}}, \bibinfo {author} {\bibfnamefont {T.}~\bibnamefont {Okane}}, \bibinfo {author} {\bibfnamefont {Y.}~\bibnamefont {Saitoh}}, \bibinfo {author} {\bibfnamefont {K.}~\bibnamefont {Kobayashi}}, \bibinfo {author} {\bibfnamefont {H.}~\bibnamefont {Yamagami}}, \bibinfo {author} {\bibfnamefont {A.}~\bibnamefont {Nakamura}}, \bibinfo {author} {\bibfnamefont {M.}~\bibnamefont {Hedo}}, \bibinfo {author} {\bibfnamefont {T.}~\bibnamefont {Nakama}},\ and\ \bibinfo {author} {\bibfnamefont {Y.}~\bibnamefont {Ōnuki}},\ }\bibfield  {title} {\bibinfo {title} {Electronic structure of \ce{EuAl4} studied by photoelectron spectroscopy},\ }\href {https://doi.org/https://doi.org/10.7566/JPSJ.85.094703} {\bibfield  {journal} {\bibinfo  {journal} {Journal of the Physical Society of Japan}\ }\textbf {\bibinfo {volume} {85}},\
  \bibinfo {pages} {094703} (\bibinfo {year} {2016})}\BibitemShut {NoStop}%
\bibitem [{\citenamefont {Shimomura}\ \emph {et~al.}(2019)\citenamefont {Shimomura}, \citenamefont {Murao}, \citenamefont {Tsutsui}, \citenamefont {Nakao}, \citenamefont {Nakamura}, \citenamefont {Hedo}, \citenamefont {Nakama},\ and\ \citenamefont {Ōnuki}}]{014602}%
  \BibitemOpen
  \bibfield  {author} {\bibinfo {author} {\bibfnamefont {S.}~\bibnamefont {Shimomura}}, \bibinfo {author} {\bibfnamefont {H.}~\bibnamefont {Murao}}, \bibinfo {author} {\bibfnamefont {S.}~\bibnamefont {Tsutsui}}, \bibinfo {author} {\bibfnamefont {H.}~\bibnamefont {Nakao}}, \bibinfo {author} {\bibfnamefont {A.}~\bibnamefont {Nakamura}}, \bibinfo {author} {\bibfnamefont {M.}~\bibnamefont {Hedo}}, \bibinfo {author} {\bibfnamefont {T.}~\bibnamefont {Nakama}},\ and\ \bibinfo {author} {\bibfnamefont {Y.}~\bibnamefont {Ōnuki}},\ }\bibfield  {title} {\bibinfo {title} {Lattice modulation and structural phase transition in the antiferromagnet \ce{EuAl4}},\ }\href {https://doi.org/https://doi.org/10.7566/JPSJ.88.014602} {\bibfield  {journal} {\bibinfo  {journal} {Journal of the Physical Society of Japan}\ }\textbf {\bibinfo {volume} {88}},\ \bibinfo {pages} {014602} (\bibinfo {year} {2019})}\BibitemShut {NoStop}%
\bibitem [{\citenamefont {Kaneko}\ \emph {et~al.}(2021)\citenamefont {Kaneko}, \citenamefont {Kawasaki}, \citenamefont {Nakamura}, \citenamefont {Munakata}, \citenamefont {Nakao}, \citenamefont {Hanashima}, \citenamefont {Kiyanagi}, \citenamefont {Ohhara}, \citenamefont {Hedo}, \citenamefont {Nakama},\ and\ \citenamefont {Ōnuki}}]{064704}%
  \BibitemOpen
  \bibfield  {author} {\bibinfo {author} {\bibfnamefont {K.}~\bibnamefont {Kaneko}}, \bibinfo {author} {\bibfnamefont {T.}~\bibnamefont {Kawasaki}}, \bibinfo {author} {\bibfnamefont {A.}~\bibnamefont {Nakamura}}, \bibinfo {author} {\bibfnamefont {K.}~\bibnamefont {Munakata}}, \bibinfo {author} {\bibfnamefont {A.}~\bibnamefont {Nakao}}, \bibinfo {author} {\bibfnamefont {T.}~\bibnamefont {Hanashima}}, \bibinfo {author} {\bibfnamefont {R.}~\bibnamefont {Kiyanagi}}, \bibinfo {author} {\bibfnamefont {T.}~\bibnamefont {Ohhara}}, \bibinfo {author} {\bibfnamefont {M.}~\bibnamefont {Hedo}}, \bibinfo {author} {\bibfnamefont {T.}~\bibnamefont {Nakama}},\ and\ \bibinfo {author} {\bibfnamefont {Y.}~\bibnamefont {Ōnuki}},\ }\bibfield  {title} {\bibinfo {title} {Charge-density-wave order and multiple magnetic transitions in divalent europium compound \ce{EuAl4}},\ }\href {https://doi.org/https://doi.org/10.7566/JPSJ.90.064704} {\bibfield  {journal} {\bibinfo  {journal} {Journal of the Physical Society of Japan}\ }\textbf
  {\bibinfo {volume} {90}},\ \bibinfo {pages} {064704} (\bibinfo {year} {2021})}\BibitemShut {NoStop}%
\bibitem [{\citenamefont {Ramakrishnan}\ \emph {et~al.}(2024)\citenamefont {Ramakrishnan}, \citenamefont {Kotla}, \citenamefont {Pi}, \citenamefont {Maity}, \citenamefont {Chen}, \citenamefont {Bao}, \citenamefont {Guo}, \citenamefont {Kado}, \citenamefont {Agarwal}, \citenamefont {Eisele}, \citenamefont {Nohara}, \citenamefont {Noohinejad}, \citenamefont {Weng}, \citenamefont {Ramakrishnan}, \citenamefont {Thamizhavel},\ and\ \citenamefont {van Smaalen}}]{023277}%
  \BibitemOpen
  \bibfield  {author} {\bibinfo {author} {\bibfnamefont {S.}~\bibnamefont {Ramakrishnan}}, \bibinfo {author} {\bibfnamefont {S.~R.}\ \bibnamefont {Kotla}}, \bibinfo {author} {\bibfnamefont {H.}~\bibnamefont {Pi}}, \bibinfo {author} {\bibfnamefont {B.~B.}\ \bibnamefont {Maity}}, \bibinfo {author} {\bibfnamefont {J.}~\bibnamefont {Chen}}, \bibinfo {author} {\bibfnamefont {J.-K.}\ \bibnamefont {Bao}}, \bibinfo {author} {\bibfnamefont {Z.}~\bibnamefont {Guo}}, \bibinfo {author} {\bibfnamefont {M.}~\bibnamefont {Kado}}, \bibinfo {author} {\bibfnamefont {H.}~\bibnamefont {Agarwal}}, \bibinfo {author} {\bibfnamefont {C.}~\bibnamefont {Eisele}}, \bibinfo {author} {\bibfnamefont {M.}~\bibnamefont {Nohara}}, \bibinfo {author} {\bibfnamefont {L.}~\bibnamefont {Noohinejad}}, \bibinfo {author} {\bibfnamefont {H.}~\bibnamefont {Weng}}, \bibinfo {author} {\bibfnamefont {S.}~\bibnamefont {Ramakrishnan}}, \bibinfo {author} {\bibfnamefont {A.}~\bibnamefont {Thamizhavel}},\ and\ \bibinfo {author} {\bibfnamefont
  {S.}~\bibnamefont {van Smaalen}},\ }\bibfield  {title} {\bibinfo {title} {Noncentrosymmetric, transverse structural modulation in \ce{SrAl4}, and elucidation of its origin in the \ce{BaAl4} family of compounds},\ }\href {https://doi.org/10.1103/PhysRevResearch.6.023277} {\bibfield  {journal} {\bibinfo  {journal} {Phys. Rev. Res.}\ }\textbf {\bibinfo {volume} {6}},\ \bibinfo {pages} {023277} (\bibinfo {year} {2024})}\BibitemShut {NoStop}%
\bibitem [{\citenamefont {Korshunov}\ \emph {et~al.}(2024)\citenamefont {Korshunov}, \citenamefont {Sukhanov}, \citenamefont {Gebel}, \citenamefont {Pavlovskii}, \citenamefont {Andriushin}, \citenamefont {Gao}, \citenamefont {Moya}, \citenamefont {Morosan},\ and\ \citenamefont {Rahn}}]{045102}%
  \BibitemOpen
  \bibfield  {author} {\bibinfo {author} {\bibfnamefont {A.~N.}\ \bibnamefont {Korshunov}}, \bibinfo {author} {\bibfnamefont {A.~S.}\ \bibnamefont {Sukhanov}}, \bibinfo {author} {\bibfnamefont {S.}~\bibnamefont {Gebel}}, \bibinfo {author} {\bibfnamefont {M.~S.}\ \bibnamefont {Pavlovskii}}, \bibinfo {author} {\bibfnamefont {N.~D.}\ \bibnamefont {Andriushin}}, \bibinfo {author} {\bibfnamefont {Y.}~\bibnamefont {Gao}}, \bibinfo {author} {\bibfnamefont {J.~M.}\ \bibnamefont {Moya}}, \bibinfo {author} {\bibfnamefont {E.}~\bibnamefont {Morosan}},\ and\ \bibinfo {author} {\bibfnamefont {M.~C.}\ \bibnamefont {Rahn}},\ }\bibfield  {title} {\bibinfo {title} {Phonon softening and atomic modulations in \ce{EuAl4}},\ }\href {https://doi.org/10.1103/PhysRevB.110.045102} {\bibfield  {journal} {\bibinfo  {journal} {Phys. Rev. B}\ }\textbf {\bibinfo {volume} {110}},\ \bibinfo {pages} {045102} (\bibinfo {year} {2024})}\BibitemShut {NoStop}%
\bibitem [{\citenamefont {Sukhanov}\ \emph {et~al.}(2025)\citenamefont {Sukhanov}, \citenamefont {Gebel}, \citenamefont {Korshunov}, \citenamefont {Andriushin}, \citenamefont {Pavlovskii}, \citenamefont {Gao}, \citenamefont {Moya}, \citenamefont {Allen}, \citenamefont {Morosan},\ and\ \citenamefont {Rahn}}]{195150}%
  \BibitemOpen
  \bibfield  {author} {\bibinfo {author} {\bibfnamefont {A.~S.}\ \bibnamefont {Sukhanov}}, \bibinfo {author} {\bibfnamefont {S.}~\bibnamefont {Gebel}}, \bibinfo {author} {\bibfnamefont {A.~N.}\ \bibnamefont {Korshunov}}, \bibinfo {author} {\bibfnamefont {N.~D.}\ \bibnamefont {Andriushin}}, \bibinfo {author} {\bibfnamefont {M.~S.}\ \bibnamefont {Pavlovskii}}, \bibinfo {author} {\bibfnamefont {Y.}~\bibnamefont {Gao}}, \bibinfo {author} {\bibfnamefont {J.~M.}\ \bibnamefont {Moya}}, \bibinfo {author} {\bibfnamefont {K.}~\bibnamefont {Allen}}, \bibinfo {author} {\bibfnamefont {E.}~\bibnamefont {Morosan}},\ and\ \bibinfo {author} {\bibfnamefont {M.~C.}\ \bibnamefont {Rahn}},\ }\bibfield  {title} {\bibinfo {title} {Electron-phonon coupling in \ce{EuAl4} under hydrostatic pressure},\ }\href {https://doi.org/10.1103/PhysRevB.111.195150} {\bibfield  {journal} {\bibinfo  {journal} {Phys. Rev. B}\ }\textbf {\bibinfo {volume} {111}},\ \bibinfo {pages} {195150} (\bibinfo {year} {2025})}\BibitemShut {NoStop}%
\bibitem [{\citenamefont {Cao}\ \emph {et~al.}(2025)\citenamefont {Cao}, \citenamefont {Jin}, \citenamefont {Zhao}, \citenamefont {Long}, \citenamefont {Luo}, \citenamefont {Zhang},\ and\ \citenamefont {Chen}}]{Cao2025}%
  \BibitemOpen
  \bibfield  {author} {\bibinfo {author} {\bibfnamefont {S.}~\bibnamefont {Cao}}, \bibinfo {author} {\bibfnamefont {F.}~\bibnamefont {Jin}}, \bibinfo {author} {\bibfnamefont {J.}~\bibnamefont {Zhao}}, \bibinfo {author} {\bibfnamefont {Y.-Z.}\ \bibnamefont {Long}}, \bibinfo {author} {\bibfnamefont {J.}~\bibnamefont {Luo}}, \bibinfo {author} {\bibfnamefont {Q.}~\bibnamefont {Zhang}},\ and\ \bibinfo {author} {\bibfnamefont {Z.-G.}\ \bibnamefont {Chen}},\ }\bibfield  {title} {\bibinfo {title} {Enhanced phonon--phonon interactions and weakened electron--phonon coupling in charge density wave topological semimetal \ce{EuAl4} with a possible intermediate electronic state},\ }\href {https://doi.org/10.1021/acs.jpclett.5c00007} {\bibfield  {journal} {\bibinfo  {journal} {The Journal of Physical Chemistry Letters}\ }\textbf {\bibinfo {volume} {16}},\ \bibinfo {pages} {1909} (\bibinfo {year} {2025})}\BibitemShut {NoStop}%
\bibitem [{\citenamefont {Wang}\ \emph {et~al.}(2024)\citenamefont {Wang}, \citenamefont {Nepal},\ and\ \citenamefont {Canfield}}]{Wang2024}%
  \BibitemOpen
  \bibfield  {author} {\bibinfo {author} {\bibfnamefont {L.-L.}\ \bibnamefont {Wang}}, \bibinfo {author} {\bibfnamefont {N.~K.}\ \bibnamefont {Nepal}},\ and\ \bibinfo {author} {\bibfnamefont {P.~C.}\ \bibnamefont {Canfield}},\ }\bibfield  {title} {\bibinfo {title} {Origin of charge density wave in topological semimetals \ce{SrAl4} and \ce{EuAl4}},\ }\href {https://doi.org/10.1038/s42005-024-01600-1} {\bibfield  {journal} {\bibinfo  {journal} {Communications Physics}\ }\textbf {\bibinfo {volume} {7}},\ \bibinfo {pages} {111} (\bibinfo {year} {2024})}\BibitemShut {NoStop}%
\bibitem [{\citenamefont {Jung}\ \emph {et~al.}(2018)\citenamefont {Jung}, \citenamefont {Park},\ and\ \citenamefont {Ihm}}]{Jung2018}%
  \BibitemOpen
  \bibfield  {author} {\bibinfo {author} {\bibfnamefont {J.~H.}\ \bibnamefont {Jung}}, \bibinfo {author} {\bibfnamefont {C.-H.}\ \bibnamefont {Park}},\ and\ \bibinfo {author} {\bibfnamefont {J.}~\bibnamefont {Ihm}},\ }\bibfield  {title} {\bibinfo {title} {A rigorous method of calculating exfoliation energies from first principles},\ }\href {https://doi.org/10.1021/acs.nanolett.7b04201} {\bibfield  {journal} {\bibinfo  {journal} {Nano Letters}\ }\textbf {\bibinfo {volume} {18}},\ \bibinfo {pages} {2759} (\bibinfo {year} {2018})}\BibitemShut {NoStop}%
\bibitem [{\citenamefont {Xiao}\ \emph {et~al.}(2024)\citenamefont {Xiao}, \citenamefont {Dong}, \citenamefont {Wang}, \citenamefont {Yu}, \citenamefont {Fu}, \citenamefont {Hu}, \citenamefont {Guo}, \citenamefont {Zhang}, \citenamefont {Hou}, \citenamefont {Guo}, \citenamefont {Yang}, \citenamefont {Xu}, \citenamefont {Tang}, \citenamefont {Duan}, \citenamefont {Xue},\ and\ \citenamefont {Li}}]{Xiao2024}%
  \BibitemOpen
  \bibfield  {author} {\bibinfo {author} {\bibfnamefont {K.}~\bibnamefont {Xiao}}, \bibinfo {author} {\bibfnamefont {W.-H.}\ \bibnamefont {Dong}}, \bibinfo {author} {\bibfnamefont {X.}~\bibnamefont {Wang}}, \bibinfo {author} {\bibfnamefont {J.}~\bibnamefont {Yu}}, \bibinfo {author} {\bibfnamefont {D.}~\bibnamefont {Fu}}, \bibinfo {author} {\bibfnamefont {Z.}~\bibnamefont {Hu}}, \bibinfo {author} {\bibfnamefont {Y.}~\bibnamefont {Guo}}, \bibinfo {author} {\bibfnamefont {Q.}~\bibnamefont {Zhang}}, \bibinfo {author} {\bibfnamefont {X.}~\bibnamefont {Hou}}, \bibinfo {author} {\bibfnamefont {Y.}~\bibnamefont {Guo}}, \bibinfo {author} {\bibfnamefont {L.}~\bibnamefont {Yang}}, \bibinfo {author} {\bibfnamefont {Y.}~\bibnamefont {Xu}}, \bibinfo {author} {\bibfnamefont {P.}~\bibnamefont {Tang}}, \bibinfo {author} {\bibfnamefont {W.}~\bibnamefont {Duan}}, \bibinfo {author} {\bibfnamefont {Q.}~\bibnamefont {Xue}},\ and\ \bibinfo {author} {\bibfnamefont {W.}~\bibnamefont {Li}},\ }\bibfield  {title} {\bibinfo {title}
  {Hidden charge order and multiple electronic instabilities in \ce{EuTe4}},\ }\href {https://doi.org/10.1021/acs.nanolett.4c01588} {\bibfield  {journal} {\bibinfo  {journal} {Nano Letters}\ }\textbf {\bibinfo {volume} {24}},\ \bibinfo {pages} {7681} (\bibinfo {year} {2024})}\BibitemShut {NoStop}%
\bibitem [{\citenamefont {Li}\ \emph {et~al.}(2021)\citenamefont {Li}, \citenamefont {Deng}, \citenamefont {Wang}, \citenamefont {Li}, \citenamefont {Liu}, \citenamefont {Zhu}, \citenamefont {Jia}, \citenamefont {Sun}, \citenamefont {Du}, \citenamefont {Yu}, \citenamefont {Guan}, \citenamefont {Wu}, \citenamefont {Zhang}, \citenamefont {Shi},\ and\ \citenamefont {Mao}}]{Li2021}%
  \BibitemOpen
  \bibfield  {author} {\bibinfo {author} {\bibfnamefont {Y.}~\bibnamefont {Li}}, \bibinfo {author} {\bibfnamefont {H.-B.}\ \bibnamefont {Deng}}, \bibinfo {author} {\bibfnamefont {C.-X.}\ \bibnamefont {Wang}}, \bibinfo {author} {\bibfnamefont {S.-S.}\ \bibnamefont {Li}}, \bibinfo {author} {\bibfnamefont {L.-M.}\ \bibnamefont {Liu}}, \bibinfo {author} {\bibfnamefont {C.-J.}\ \bibnamefont {Zhu}}, \bibinfo {author} {\bibfnamefont {K.}~\bibnamefont {Jia}}, \bibinfo {author} {\bibfnamefont {Y.-K.}\ \bibnamefont {Sun}}, \bibinfo {author} {\bibfnamefont {X.}~\bibnamefont {Du}}, \bibinfo {author} {\bibfnamefont {X.}~\bibnamefont {Yu}}, \bibinfo {author} {\bibfnamefont {T.}~\bibnamefont {Guan}}, \bibinfo {author} {\bibfnamefont {R.}~\bibnamefont {Wu}}, \bibinfo {author} {\bibfnamefont {S.-Y.}\ \bibnamefont {Zhang}}, \bibinfo {author} {\bibfnamefont {Y.-G.}\ \bibnamefont {Shi}},\ and\ \bibinfo {author} {\bibfnamefont {H.-Q.}\ \bibnamefont {Mao}},\ }\bibfield  {title} {\bibinfo {title} {Surface and electronic structure
  of antiferromagnetic axion insulator candidate \ce{EuIn2As2}},\ }\href {https://doi.org/10.7498/aps.70.20210783} {\bibfield  {journal} {\bibinfo  {journal} {Acta Physica Sinica}\ }\textbf {\bibinfo {volume} {70}},\ \bibinfo {pages} {186801} (\bibinfo {year} {2021})}\BibitemShut {NoStop}%
\bibitem [{\citenamefont {Burk}\ \emph {et~al.}(1992)\citenamefont {Burk}, \citenamefont {Thomson}, \citenamefont {Clarke},\ and\ \citenamefont {Zettl}}]{5068362}%
  \BibitemOpen
  \bibfield  {author} {\bibinfo {author} {\bibfnamefont {B.}~\bibnamefont {Burk}}, \bibinfo {author} {\bibfnamefont {R.~E.}\ \bibnamefont {Thomson}}, \bibinfo {author} {\bibfnamefont {J.}~\bibnamefont {Clarke}},\ and\ \bibinfo {author} {\bibfnamefont {A.}~\bibnamefont {Zettl}},\ }\bibfield  {title} {\bibinfo {title} {Surface and bulk charge density wave structure in 1 \ce{T}-\ce{TaS2}},\ }\href {https://doi.org/10.1126/science.257.5068.362} {\bibfield  {journal} {\bibinfo  {journal} {Science}\ }\textbf {\bibinfo {volume} {257}},\ \bibinfo {pages} {362} (\bibinfo {year} {1992})}\BibitemShut {NoStop}%
\bibitem [{\citenamefont {Fu}\ \emph {et~al.}(2016)\citenamefont {Fu}, \citenamefont {Kraft}, \citenamefont {Sharma}, \citenamefont {Singh}, \citenamefont {Walmsley}, \citenamefont {Fisher},\ and\ \citenamefont {Boyer}}]{205101}%
  \BibitemOpen
  \bibfield  {author} {\bibinfo {author} {\bibfnamefont {L.}~\bibnamefont {Fu}}, \bibinfo {author} {\bibfnamefont {A.~M.}\ \bibnamefont {Kraft}}, \bibinfo {author} {\bibfnamefont {B.}~\bibnamefont {Sharma}}, \bibinfo {author} {\bibfnamefont {M.}~\bibnamefont {Singh}}, \bibinfo {author} {\bibfnamefont {P.}~\bibnamefont {Walmsley}}, \bibinfo {author} {\bibfnamefont {I.~R.}\ \bibnamefont {Fisher}},\ and\ \bibinfo {author} {\bibfnamefont {M.~C.}\ \bibnamefont {Boyer}},\ }\bibfield  {title} {\bibinfo {title} {Multiple charge density wave states at the surface of $\mathrm{TbT}{\mathrm{e}}_{3}$},\ }\href {https://doi.org/10.1103/PhysRevB.94.205101} {\bibfield  {journal} {\bibinfo  {journal} {Phys. Rev. B}\ }\textbf {\bibinfo {volume} {94}},\ \bibinfo {pages} {205101} (\bibinfo {year} {2016})}\BibitemShut {NoStop}%
\bibitem [{\citenamefont {Chen}\ \emph {et~al.}(2019)\citenamefont {Chen}, \citenamefont {Wang}, \citenamefont {Wu}, \citenamefont {Ma}, \citenamefont {Wen}, \citenamefont {Wu}, \citenamefont {Li}, \citenamefont {Zhao}, \citenamefont {Wang}, \citenamefont {Zhang}, \citenamefont {Huang}, \citenamefont {Li},\ and\ \citenamefont {Huang}}]{5118870}%
  \BibitemOpen
  \bibfield  {author} {\bibinfo {author} {\bibfnamefont {Y.}~\bibnamefont {Chen}}, \bibinfo {author} {\bibfnamefont {P.}~\bibnamefont {Wang}}, \bibinfo {author} {\bibfnamefont {M.}~\bibnamefont {Wu}}, \bibinfo {author} {\bibfnamefont {J.}~\bibnamefont {Ma}}, \bibinfo {author} {\bibfnamefont {S.}~\bibnamefont {Wen}}, \bibinfo {author} {\bibfnamefont {X.}~\bibnamefont {Wu}}, \bibinfo {author} {\bibfnamefont {G.}~\bibnamefont {Li}}, \bibinfo {author} {\bibfnamefont {Y.}~\bibnamefont {Zhao}}, \bibinfo {author} {\bibfnamefont {K.}~\bibnamefont {Wang}}, \bibinfo {author} {\bibfnamefont {L.}~\bibnamefont {Zhang}}, \bibinfo {author} {\bibfnamefont {L.}~\bibnamefont {Huang}}, \bibinfo {author} {\bibfnamefont {W.}~\bibnamefont {Li}},\ and\ \bibinfo {author} {\bibfnamefont {M.}~\bibnamefont {Huang}},\ }\bibfield  {title} {\bibinfo {title} {Raman spectra and dimensional effect on the charge density wave transition in \ce{GdTe3}},\ }\href {https://doi.org/10.1063/1.5118870} {\bibfield  {journal} {\bibinfo  {journal}
  {Applied Physics Letters}\ }\textbf {\bibinfo {volume} {115}},\ \bibinfo {pages} {151905} (\bibinfo {year} {2019})}\BibitemShut {NoStop}%
\bibitem [{\citenamefont {Liu}\ \emph {et~al.}(2024)\citenamefont {Liu}, \citenamefont {Li}, \citenamefont {Zhu}, \citenamefont {Shou}, \citenamefont {Adam}, \citenamefont {Cui}, \citenamefont {Li}, \citenamefont {Wang}, \citenamefont {Wu}, \citenamefont {Zhu}, \citenamefont {Liu}, \citenamefont {Chen}, \citenamefont {Wu}, \citenamefont {Cui}, \citenamefont {Song},\ and\ \citenamefont {Sun}}]{Liu2024}%
  \BibitemOpen
  \bibfield  {author} {\bibinfo {author} {\bibfnamefont {Z.}~\bibnamefont {Liu}}, \bibinfo {author} {\bibfnamefont {T.}~\bibnamefont {Li}}, \bibinfo {author} {\bibfnamefont {W.}~\bibnamefont {Zhu}}, \bibinfo {author} {\bibfnamefont {H.}~\bibnamefont {Shou}}, \bibinfo {author} {\bibfnamefont {M.~L.}\ \bibnamefont {Adam}}, \bibinfo {author} {\bibfnamefont {Q.}~\bibnamefont {Cui}}, \bibinfo {author} {\bibfnamefont {Y.}~\bibnamefont {Li}}, \bibinfo {author} {\bibfnamefont {S.}~\bibnamefont {Wang}}, \bibinfo {author} {\bibfnamefont {Y.}~\bibnamefont {Wu}}, \bibinfo {author} {\bibfnamefont {H.}~\bibnamefont {Zhu}}, \bibinfo {author} {\bibfnamefont {Y.}~\bibnamefont {Liu}}, \bibinfo {author} {\bibfnamefont {S.}~\bibnamefont {Chen}}, \bibinfo {author} {\bibfnamefont {X.}~\bibnamefont {Wu}}, \bibinfo {author} {\bibfnamefont {S.}~\bibnamefont {Cui}}, \bibinfo {author} {\bibfnamefont {L.}~\bibnamefont {Song}},\ and\ \bibinfo {author} {\bibfnamefont {Z.}~\bibnamefont {Sun}},\ }\bibfield  {title} {\bibinfo {title}
  {Persistence of charge density wave against variation of band structures in \ce{V_{x}Ti_{1$-$x}Se2(x = $0-0.1$)}},\ }\href {https://doi.org/10.1007/s12274-023-5936-z} {\bibfield  {journal} {\bibinfo  {journal} {Nano Research}\ }\textbf {\bibinfo {volume} {17}},\ \bibinfo {pages} {2129} (\bibinfo {year} {2024})}\BibitemShut {NoStop}%
\bibitem [{\citenamefont {Zhu}\ \emph {et~al.}(2015)\citenamefont {Zhu}, \citenamefont {Cao}, \citenamefont {Zhang}, \citenamefont {Plummer},\ and\ \citenamefont {Guo}}]{1424791112}%
  \BibitemOpen
  \bibfield  {author} {\bibinfo {author} {\bibfnamefont {X.}~\bibnamefont {Zhu}}, \bibinfo {author} {\bibfnamefont {Y.}~\bibnamefont {Cao}}, \bibinfo {author} {\bibfnamefont {J.}~\bibnamefont {Zhang}}, \bibinfo {author} {\bibfnamefont {E.~W.}\ \bibnamefont {Plummer}},\ and\ \bibinfo {author} {\bibfnamefont {J.}~\bibnamefont {Guo}},\ }\bibfield  {title} {\bibinfo {title} {Classification of charge density waves based on their nature},\ }\href {https://doi.org/10.1073/pnas.1424791112} {\bibfield  {journal} {\bibinfo  {journal} {Proceedings of the National Academy of Sciences}\ }\textbf {\bibinfo {volume} {112}},\ \bibinfo {pages} {2367} (\bibinfo {year} {2015})}\BibitemShut {NoStop}%
\bibitem [{\citenamefont {Chen}\ \emph {et~al.}(2015)\citenamefont {Chen}, \citenamefont {Chan}, \citenamefont {Fang}, \citenamefont {Zhang}, \citenamefont {Chou}, \citenamefont {Mo}, \citenamefont {Hussain}, \citenamefont {Fedorov},\ and\ \citenamefont {Chiang}}]{Chen2015}%
  \BibitemOpen
  \bibfield  {author} {\bibinfo {author} {\bibfnamefont {P.}~\bibnamefont {Chen}}, \bibinfo {author} {\bibfnamefont {Y.-H.}\ \bibnamefont {Chan}}, \bibinfo {author} {\bibfnamefont {X.-Y.}\ \bibnamefont {Fang}}, \bibinfo {author} {\bibfnamefont {Y.}~\bibnamefont {Zhang}}, \bibinfo {author} {\bibfnamefont {M.~Y.}\ \bibnamefont {Chou}}, \bibinfo {author} {\bibfnamefont {S.-K.}\ \bibnamefont {Mo}}, \bibinfo {author} {\bibfnamefont {Z.}~\bibnamefont {Hussain}}, \bibinfo {author} {\bibfnamefont {A.-V.}\ \bibnamefont {Fedorov}},\ and\ \bibinfo {author} {\bibfnamefont {T.-C.}\ \bibnamefont {Chiang}},\ }\bibfield  {title} {\bibinfo {title} {Charge density wave transition in single-layer titanium diselenide},\ }\href {https://doi.org/10.1038/ncomms9943} {\bibfield  {journal} {\bibinfo  {journal} {Nature Communications}\ }\textbf {\bibinfo {volume} {6}},\ \bibinfo {pages} {8943} (\bibinfo {year} {2015})}\BibitemShut {NoStop}%
\bibitem [{\citenamefont {Yang}\ \emph {et~al.}(2021)\citenamefont {Yang}, \citenamefont {Liu}, \citenamefont {Liu}, \citenamefont {Liu}, \citenamefont {Liu}, \citenamefont {Lu}, \citenamefont {Mei}, \citenamefont {Li}, \citenamefont {Ye}, \citenamefont {Qiao} \emph {et~al.}}]{yang2021high}%
  \BibitemOpen
  \bibfield  {author} {\bibinfo {author} {\bibfnamefont {Y.-C.}\ \bibnamefont {Yang}}, \bibinfo {author} {\bibfnamefont {Z.-T.}\ \bibnamefont {Liu}}, \bibinfo {author} {\bibfnamefont {J.-S.}\ \bibnamefont {Liu}}, \bibinfo {author} {\bibfnamefont {Z.-H.}\ \bibnamefont {Liu}}, \bibinfo {author} {\bibfnamefont {W.-L.}\ \bibnamefont {Liu}}, \bibinfo {author} {\bibfnamefont {X.-L.}\ \bibnamefont {Lu}}, \bibinfo {author} {\bibfnamefont {H.-P.}\ \bibnamefont {Mei}}, \bibinfo {author} {\bibfnamefont {A.}~\bibnamefont {Li}}, \bibinfo {author} {\bibfnamefont {M.}~\bibnamefont {Ye}}, \bibinfo {author} {\bibfnamefont {S.}~\bibnamefont {Qiao}}, \emph {et~al.},\ }\bibfield  {title} {\bibinfo {title} {High-resolution arpes endstation for in situ electronic structure investigations at \ce{SSRF}},\ }\href {https://doi.org/10.1007/s41365-021-00858-2} {\bibfield  {journal} {\bibinfo  {journal} {Nuclear Science and Techniques}\ }\textbf {\bibinfo {volume} {32}},\ \bibinfo {pages} {31} (\bibinfo {year} {2021})}\BibitemShut
  {NoStop}%
\bibitem [{\citenamefont {Sun}\ \emph {et~al.}(2020)\citenamefont {Sun}, \citenamefont {Liu}, \citenamefont {Liu}, \citenamefont {Liu}, \citenamefont {Zhang}, \citenamefont {Shen}, \citenamefont {Ye},\ and\ \citenamefont {Qiao}}]{sun2020performance}%
  \BibitemOpen
  \bibfield  {author} {\bibinfo {author} {\bibfnamefont {Z.}~\bibnamefont {Sun}}, \bibinfo {author} {\bibfnamefont {Z.}~\bibnamefont {Liu}}, \bibinfo {author} {\bibfnamefont {Z.}~\bibnamefont {Liu}}, \bibinfo {author} {\bibfnamefont {W.}~\bibnamefont {Liu}}, \bibinfo {author} {\bibfnamefont {F.}~\bibnamefont {Zhang}}, \bibinfo {author} {\bibfnamefont {D.}~\bibnamefont {Shen}}, \bibinfo {author} {\bibfnamefont {M.}~\bibnamefont {Ye}},\ and\ \bibinfo {author} {\bibfnamefont {S.}~\bibnamefont {Qiao}},\ }\bibfield  {title} {\bibinfo {title} {Performance of the \ce{BL03U} beamline at \ce{SSRF}},\ }\href {https://doi.org/https://doi.org/10.1107/S1600577520008310} {\bibfield  {journal} {\bibinfo  {journal} {Journal of Synchrotron Radiation}\ }\textbf {\bibinfo {volume} {27}},\ \bibinfo {pages} {1388} (\bibinfo {year} {2020})}\BibitemShut {NoStop}%
\bibitem [{\citenamefont {Bl\"ochl}(1994)}]{prb5017}%
  \BibitemOpen
  \bibfield  {author} {\bibinfo {author} {\bibfnamefont {P.~E.}\ \bibnamefont {Bl\"ochl}},\ }\bibfield  {title} {\bibinfo {title} {Projector augmented-wave method},\ }\href {https://doi.org/10.1103/PhysRevB.50.17953} {\bibfield  {journal} {\bibinfo  {journal} {Phys. Rev. B}\ }\textbf {\bibinfo {volume} {50}},\ \bibinfo {pages} {17953} (\bibinfo {year} {1994})}\BibitemShut {NoStop}%
\bibitem [{\citenamefont {Kresse}\ and\ \citenamefont {Furthm\"uller}(1996)}]{prb511169}%
  \BibitemOpen
  \bibfield  {author} {\bibinfo {author} {\bibfnamefont {G.}~\bibnamefont {Kresse}}\ and\ \bibinfo {author} {\bibfnamefont {J.}~\bibnamefont {Furthm\"uller}},\ }\bibfield  {title} {\bibinfo {title} {Efficient iterative schemes for ab initio total-energy calculations using a plane-wave basis set},\ }\href {https://doi.org/10.1103/PhysRevB.54.11169} {\bibfield  {journal} {\bibinfo  {journal} {Phys. Rev. B}\ }\textbf {\bibinfo {volume} {54}},\ \bibinfo {pages} {11169} (\bibinfo {year} {1996})}\BibitemShut {NoStop}%
\bibitem [{\citenamefont {Perdew}\ \emph {et~al.}(1996)\citenamefont {Perdew}, \citenamefont {Burke},\ and\ \citenamefont {Ernzerhof}}]{prl773865}%
  \BibitemOpen
  \bibfield  {author} {\bibinfo {author} {\bibfnamefont {J.~P.}\ \bibnamefont {Perdew}}, \bibinfo {author} {\bibfnamefont {K.}~\bibnamefont {Burke}},\ and\ \bibinfo {author} {\bibfnamefont {M.}~\bibnamefont {Ernzerhof}},\ }\bibfield  {title} {\bibinfo {title} {Generalized gradient approximation made simple},\ }\href {https://doi.org/10.1103/PhysRevLett.77.3865} {\bibfield  {journal} {\bibinfo  {journal} {Phys. Rev. Lett.}\ }\textbf {\bibinfo {volume} {77}},\ \bibinfo {pages} {3865} (\bibinfo {year} {1996})}\BibitemShut {NoStop}%
\bibitem [{\citenamefont {Marzari}\ and\ \citenamefont {Vanderbilt}(1997)}]{prb5612847}%
  \BibitemOpen
  \bibfield  {author} {\bibinfo {author} {\bibfnamefont {N.}~\bibnamefont {Marzari}}\ and\ \bibinfo {author} {\bibfnamefont {D.}~\bibnamefont {Vanderbilt}},\ }\bibfield  {title} {\bibinfo {title} {Maximally localized generalized wannier functions for composite energy bands},\ }\href {https://doi.org/10.1103/PhysRevB.56.12847} {\bibfield  {journal} {\bibinfo  {journal} {Phys. Rev. B}\ }\textbf {\bibinfo {volume} {56}},\ \bibinfo {pages} {12847} (\bibinfo {year} {1997})}\BibitemShut {NoStop}%
\bibitem [{\citenamefont {Souza}\ \emph {et~al.}(2001)\citenamefont {Souza}, \citenamefont {Marzari},\ and\ \citenamefont {Vanderbilt}}]{prb65035109}%
  \BibitemOpen
  \bibfield  {author} {\bibinfo {author} {\bibfnamefont {I.}~\bibnamefont {Souza}}, \bibinfo {author} {\bibfnamefont {N.}~\bibnamefont {Marzari}},\ and\ \bibinfo {author} {\bibfnamefont {D.}~\bibnamefont {Vanderbilt}},\ }\bibfield  {title} {\bibinfo {title} {Maximally localized wannier functions for entangled energy bands},\ }\href {https://doi.org/10.1103/PhysRevB.65.035109} {\bibfield  {journal} {\bibinfo  {journal} {Phys. Rev. B}\ }\textbf {\bibinfo {volume} {65}},\ \bibinfo {pages} {035109} (\bibinfo {year} {2001})}\BibitemShut {NoStop}%
\bibitem [{\citenamefont {Sancho}\ \emph {et~al.}(1984)\citenamefont {Sancho}, \citenamefont {Sancho},\ and\ \citenamefont {Rubio}}]{mpls}%
  \BibitemOpen
  \bibfield  {author} {\bibinfo {author} {\bibfnamefont {M.~P.~L.}\ \bibnamefont {Sancho}}, \bibinfo {author} {\bibfnamefont {J.~M.~L.}\ \bibnamefont {Sancho}},\ and\ \bibinfo {author} {\bibfnamefont {J.}~\bibnamefont {Rubio}},\ }\bibfield  {title} {\bibinfo {title} {Quick iterative scheme for the calculation of transfer matrices: application to \ce{Mo} (100)},\ }\href {https://doi.org/10.1088/0305-4608/14/5/016} {\bibfield  {journal} {\bibinfo  {journal} {Journal of Physics F: Metal Physics}\ }\textbf {\bibinfo {volume} {14}},\ \bibinfo {pages} {1205} (\bibinfo {year} {1984})}\BibitemShut {NoStop}%
\bibitem [{\citenamefont {Sancho}\ \emph {et~al.}(1985)\citenamefont {Sancho}, \citenamefont {Sancho}, \citenamefont {Sancho},\ and\ \citenamefont {Rubio}}]{mpls2}%
  \BibitemOpen
  \bibfield  {author} {\bibinfo {author} {\bibfnamefont {M.~P.~L.}\ \bibnamefont {Sancho}}, \bibinfo {author} {\bibfnamefont {J.~M.~L.}\ \bibnamefont {Sancho}}, \bibinfo {author} {\bibfnamefont {J.~M.~L.}\ \bibnamefont {Sancho}},\ and\ \bibinfo {author} {\bibfnamefont {J.}~\bibnamefont {Rubio}},\ }\bibfield  {title} {\bibinfo {title} {Highly convergent schemes for the calculation of bulk and surface green functions},\ }\href {https://doi.org/10.1088/0305-4608/15/4/009} {\bibfield  {journal} {\bibinfo  {journal} {Journal of Physics F: Metal Physics}\ }\textbf {\bibinfo {volume} {15}},\ \bibinfo {pages} {851} (\bibinfo {year} {1985})}\BibitemShut {NoStop}%
\bibitem [{\citenamefont {Wu}\ \emph {et~al.}(2017{\natexlab{b}})\citenamefont {Wu}, \citenamefont {Zhang}, \citenamefont {Song}, \citenamefont {Troyer},\ and\ \citenamefont {Soluyanov}}]{wuwt}%
  \BibitemOpen
  \bibfield  {author} {\bibinfo {author} {\bibfnamefont {Q.}~\bibnamefont {Wu}}, \bibinfo {author} {\bibfnamefont {S.}~\bibnamefont {Zhang}}, \bibinfo {author} {\bibfnamefont {H.}~\bibnamefont {Song}}, \bibinfo {author} {\bibfnamefont {M.}~\bibnamefont {Troyer}},\ and\ \bibinfo {author} {\bibfnamefont {A.~A.}\ \bibnamefont {Soluyanov}},\ }\bibfield  {title} {\bibinfo {title} {Wanniertools: An open-source software package for novel topological materials},\ }\href {https://doi.org/https://doi.org/10.1016/j.cpc.2017.09.033} {\bibfield  {journal} {\bibinfo  {journal} {Comput. Phys. Commun.}\ }\textbf {\bibinfo {volume} {224}},\ \bibinfo {pages} {405} (\bibinfo {year} {2017}{\natexlab{b}})}\BibitemShut {NoStop}%
\end{thebibliography}%

\end{document}